\documentclass[pra,letterpaper,nobalancelastpage,twocolumn,superscriptaddress,nofootinbib,aps,floatfix,nobibnotes]{revtex4-2}
\usepackage{relsize}
\usepackage{soul}
\RequirePackage{import}
\usepackage{textcomp}  
\usepackage{tablefootnote}
\usepackage{xr-hyper}
\usepackage{graphicx}
\usepackage{amsmath}
\usepackage{lmodern}

\usepackage{bbold}
\usepackage{amssymb}
\usepackage{amsmath}
\newcommand{\wet}{Wigner--Eckart theorem}
\usepackage[english]{babel}
\usepackage{physics}
\usepackage{color}
\usepackage{booktabs}
\usepackage{bm}
\usepackage{physunits}
\usepackage[unicode]{hyperref}
\usepackage[dvipsnames]{xcolor}
\usepackage{chngcntr}
\usepackage{dsfont}
\usepackage{placeins}
\usepackage{mathtools}
\usepackage[normalem]{ulem}
\usepackage{varwidth}
\usepackage{siunitx}
\usepackage{gensymb}
\usepackage{bibunits}
\usepackage{natbib}

\usepackage{multirow}
\usepackage{booktabs}      
\usepackage{tabularx}      

\usepackage{siunitx}
\usepackage{microtype}

\usepackage{amsmath}
\usepackage{braket}
\usepackage{relsize}

\usepackage{soul}
\newcommand{\mref}[2]{%
    \hyperref[{#1}]{%
        \ref*{#1}#2%
    }%
}
\usepackage{cleveref}
\DeclarePairedDelimiterX\cusket[1]{\vert}{\rangle}{#1}
\DeclarePairedDelimiterX\cusbra[1]{\langle}{\vert}{#1}
\DeclarePairedDelimiterX\cusbraket[2]{\langle}{\rangle}{#1 \delimsize\vert #2}

\usepackage{tikz}

\usepackage{adjustbox}

\usepackage{blindtext}
\usepackage [autostyle, english = american]{csquotes}
\MakeOuterQuote{"}
\usetikzlibrary{arrows}
\usetikzlibrary{shapes.callouts}
\usetikzlibrary{calc,arrows,arrows.meta,decorations.pathmorphing,intersections}

\usetikzlibrary{decorations.pathreplacing,angles,quotes}
\tikzset{
    level/.style = {
        line width=1pt,
    },
    connect/.style = {
        line width=1pt,<->,shorten >=2pt,shorten <=2pt,>=stealth
    },
    laser/.style = {
        line width=1pt,->,shorten >=2pt,shorten <=2pt,>=stealth
    },
    notice/.style = {
        draw,
        rectangle callout,
        callout relative pointer={#1}
    },
    label/.style = {
        text width=2cm
    },
    virtual/.style={line width=1pt,densely dashed,color=gray},
    sublevel/.style={black,densely dashed},
    ionization/.style={black,dashed},
    allarrows/.style={line width=1pt,->,>=stealth',shorten >=1pt},
    absorption/.style={allarrows, yellow, >=Triangle},
    radiative/.style={allarrows,red, Triangle[blue]-Circle, decorate,decoration={snake,amplitude=1.5}},
    indirectradiative/.style={radiative,densely dashed},
    decay/.style={allarrows,gray,thin,decorate,decoration={snake,amplitude=1.5}},
}
\usepackage{orcidlink}
\newcommand{\LANL}{Experimental Quantum Group, Materials Physics and Applications Division, Los Alamos National Laboratory, Los Alamos, New Mexico 87545}
\newcommand{\JILA}{JILA, National Institute of Standards and Technology and University of Colorado, Department of Physics, Boulder, Colorado 80309, USA}
\newcommand{\UNM}{Center for Quantum Information and Control (CQuIC), Department of Physics and Astronomy, University of New Mexico, Albuquerque, New Mexico 87131, USA}

\newcommand{\groundStatefull}{$5\mathrm{s}^{2} \; {}^{1}\mathrm{S}_0$}
\newcommand{\groundState}{${}^{1}\mathrm{S}_0$\xspace}
\newcommand{\clockStatefull}{$5\mathrm{s}5\mathrm{p} \; {}^{3}\mathrm{P}_0$\xspace}
\newcommand{\clockState}{$\mathrm{^{3}P_0}$\xspace}
\newcommand{\redState}{$\mathrm{^{3}P_1}$\xspace}
\newcommand{\greenState}{$\mathrm{^{3}P_2}$\xspace}
\newcommand{\greenStatefull}{$5s5p \ \mathrm{^{3}P_2}$\xspace}

\newcommand{\metastablemanifold}{$5s5p \ \mathrm{^{3}P_J}$\xspace}
\newcommand{\bluemotleak}{$5s4d \ \mathrm{^{1}D_2}$\xspace}

\newcommand{\Sr}{${}^{87}$Sr\xspace}

\usepackage[caption=false]{subfig} 
\usepackage{ragged2e} 
\DeclareCaptionJustification{justified}{\justifying}

\usepackage{float}
\usepackage{graphicx}
\graphicspath{{figs/}}
\DeclareCaptionLabelSeparator{bar}{ \textbf{\textbar}~}
\usepackage{enumitem}
\setlist{nolistsep}

\let\oldchi\chi
\renewcommand{\chi}{%
  \raisebox{0.44ex}{$\oldchi$}%
}

\makeatletter
\newcommand*{\addFileDependency}[1]{
\typeout{(#1)}
\@addtofilelist{#1}
\IfFileExists{#1}{}{\typeout{No file #1.}}
}\makeatother

\definecolor{lime}{HTML}{A6CE39}
\DeclareRobustCommand{\orcidicon}{
	\begin{tikzpicture}
	\draw[lime, fill=lime] (0,0) 
	circle [radius=0.16] 
	node[white] {{\fontfamily{qag}\selectfont \tiny ID}};
	\draw[white, fill=white] (-0.0625,0.095) 
	circle [radius=0.007];
	\end{tikzpicture}
	\hspace{-0.3mm}
}

\foreach \x in {A, ..., Z}{\expandafter\xdef\csname orcid\x\endcsname{\noexpand\href{https://orcid.org/\csname orcidauthor\x\endcsname}
			{\noexpand\orcidicon}}
}

\usepackage{IEEEtrantools}

\def\hhexcmyk(#1,#2,#3,#4)#5{%
    \pgfmathsetmacro{\myc}{#1/255}%
    \pgfmathsetmacro{\mym}{#2/255}%
    \pgfmathsetmacro{\myy}{#3/255}%
    \pgfmathsetmacro{\myk}{#4/255}%
    \definecolor{testcolor}{cmyk}{\myc,\mym,\myy,\myk}%
    \textcolor{testcolor}{#5}
}

\def\defineCMYKcolor(#1,#2,#3,#4)#5{%
    \pgfmathsetmacro{\myc}{#1/255}%
    \pgfmathsetmacro{\mym}{#2/255}%
    \pgfmathsetmacro{\myy}{#3/255}%
    \pgfmathsetmacro{\myk}{#4/255}%
    \definecolor{#5}{cmyk}{\myc,\mym,\myy,\myk}%
}
\definecolor{bluemotcolor}{RGB}{0,127,255}
\definecolor{awesome}{rgb}{1.0, 0.13, 0.32}
\definecolor{dodgerblue}{rgb}{0.12, 0.56, 1.0}

\hypersetup{
    colorlinks=true,
    citecolor=dodgerblue,
    linkcolor=dodgerblue,
    urlcolor=dodgerblue
}

\let\oldchi\chi
\renewcommand{\chi}{%
  \raisebox{0.44ex}{$\oldchi$}%
}

\begin{document}
\font\boldface=cmb10
\font\normalweight=cmr10
\colorlet{ochre}{blue!30!yellow!70!}
\definecolor{bluemot}{rgb}{0,.42,1}
\definecolor{clock}{RGB}{255,0,0}
\definecolor{bluerepump}{rgb}{0,.82,1}
\definecolor{secondrydberglaser}{rgb}{1,.77,0}
\definecolor{rydbgerglaser}{rgb}{0.38,0,.4}

\title{Bichromatic Tweezers for Qudit Quantum Computing in \Sr}
\author{Enrique A. Segura Carrillo$\orcidA{}$}
\email{enrique.segura@colorado.edu}
\affiliation{\JILA}
\affiliation{\LANL}

\author{Eric J. Meier$\orcidB{}$}
\affiliation{\LANL}

\author{Michael J. Martin$\orcidC{}$}
\email{mmartin@lanl.gov}
\affiliation{\LANL}
\affiliation{\UNM}



\begin{abstract}
\noindent Neutral atoms have become a competitive platform for quantum metrology, simulation, sensing, and computing. Current magic trapping techniques are insufficient to engineer magic trapping conditions for qudits encoded in hyperfine states with $J\neq 0$, compromising qudit coherence. In this paper we propose a scheme to engineer magic trapping conditions for qudits via bichromatic tweezers. We show it is possible to suppress differential light shifts across all magnetic sublevels of the \greenStatefull state by using two carefully chosen wavelengths (with comparable tensor light shift magnitude and opposite sign) at an appropriate intensity ratio, thus suppressing light-shift induced dephasing, enabling scalar magic conditions between the ground state and \greenStatefull, and tensor magic conditions for qudits encoded within it. Furthermore, this technique enables robust operation at the tensor magic angle 54.7$^\circ$ with linear trap polarization via reduced sensitivity to uncertainty in experimental parameters. We expect this technique to enable new loading protocols, enhance cooling efficiency, and enhance nuclear spins’ coherence times, thus facilitating qudit-based quantum computing in \Sr in the \greenStatefull manifold.
\end{abstract}
\maketitle

\section{Introduction}
\begin{figure}[!h]
    \includegraphics[keepaspectratio]{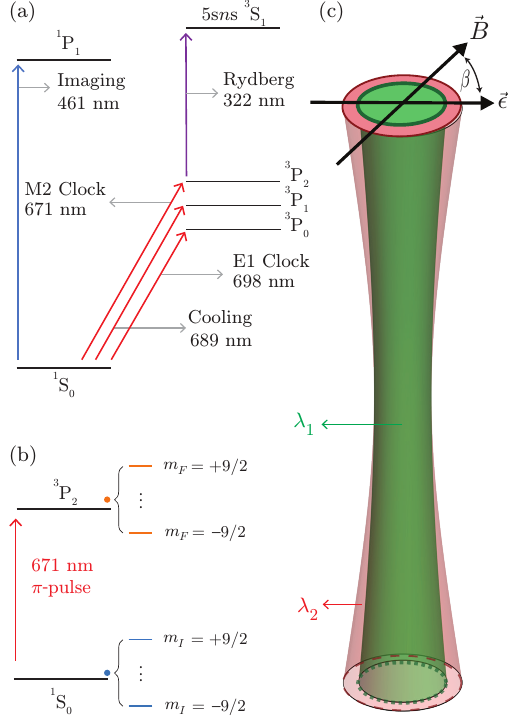}
        \caption{ \textbf{Light Shift Engineering in \Sr via Bichromatic Tweezer.} (a) Relevant Energy Levels in \Sr for this work. (b) Representation of nuclear spins in \Sr. The nuclear spin $I=9/2$ introduces 10 states in \groundState comprising a qudit. Through a coherent, linearly polarized, excitation at 671\,nm (red line), an atom is excited to \greenStatefull for quantum operations. (c) Experimental concept of Bichromatic tweezers. By aligning the orientation of the quantization axis, defined by the magnetic field $\vec{\mathbf{B}}$, relative to the polarization vector $\epsilon$ by angle $\beta$, differential light shifts between two atomic states can be manipulated.} 
    \label{fig:overview}
\end{figure}

Neutral atoms are among the leading candidates for  quantum simulation, sensing, and computing because of their long coherence times, scalability, and robustness~\cite{ni_adam,Barnes_2022}. Alkaline-Earth atoms further enhance these features by introducing narrow~\cite{katori_3p1,Yamamoto_magic_yb} and ultranarrow~\cite{jun_rdme, jun_eit,klusener} lines (see Fig.~\ref{fig:overview}a), which have been leveraged in narrow-line cooling schemes yielding atomic ensembles at ultra-cold temperatures~\cite{caltech_tweezers,katori_3p1_cooling,jeff_magic_yb,urech,benspaper}. Among alkaline-earth atoms, fermionic strontium, with $I=9/2$ and dimension $d=10$, possesses the largest nuclear spin manifold~\cite{jun_symmetry_structure} (Fig.~\ref{fig:overview}b). Nuclear spins in \Sr enable the development of a new qudit-based quantum computing architecture~\cite{Barnes_2022,Omanakuttan_2021,jun_state_dependent_lattice} with long coherence times~\cite{Barnes_2022, CatStateMinuteCoherence}, and opportunity for error correcting protocols~\cite{siva_qec}.

Optical tweezers' usefulness in assembling large atomic arrays while retaining individual atom control has made them the building blocks of neutral atom quantum computers and tweezer atomic clocks~\cite{hannah_tweezers, aaron_half_minute_tc, ivaylo_813_clock,enders_813,Cao_813_tweezer_clock,Finkelstein_813_tweezer_clock_progammable}. However, there is a trade-off: optical tweezers introduce decoherence and dephasing mechanisms~\cite{arno_dephasing}. Decoherence appears via Raman and Rayleigh scattering~\cite{lisdat_scattering, era_uys}, inducing spontaneous optical pumping, leading to leakage errors, limiting qudit coherence~\cite{siva_qec}. Dephasing emerges from nuclear-spin dependent differential light shifts, leading to gate infidelity~\cite{era_joanna}. 

Light shift engineering is crucial for exploiting the hyperfine degrees of freedom in the ground and excited electronic states in \Sr~\cite{siva_rydbgerg,Omanakuttan_2021}. Furthermore, it is a key component in developing information-preserving cooling schemes~\cite{cooling_ivan,cooling_weak,siva_qnd_cooling}, engineering new high-fidelity entangling gates~\cite{siva_rydbgerg}, implementing nuclear spin state preparation, manipulation and detection protocols~\cite{osg}, creating novel state-selective potentials~\cite{yb_state_dependent_potentials,jun_state_dependent_lattice}, and assembling large, scalable, controllable, atomic arrays via optical tweezers~\cite{caltech_tweezers,hannah_tweezers}.

\begin{figure}[!h]
    \includegraphics{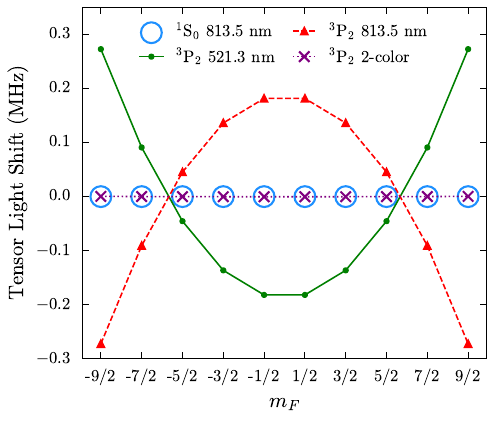}
        \caption{ \textbf{Flattening of the Tensor Manifold in \greenState} We illustrate the cancellation of tensor light shift induced by 813.5\,nm for a laser power of 1 \unit{\mW} and a beam waist $w_0$ = 1.0 microns. The red dashed line with triangle markers represents the tensor light shift induced at 813.5\,nm. We calculate a quadratic light shift dependency on nuclear spin $m_F$, which is on the scale of 0.3 \unit{\mega\hertz}. The green solid line with filled circular markers represents the light shift induced in \greenState at 521\,nm. The blue unfilled circles represent the light shift present in \groundState at 813\,nm. The purple line with cross markers represents the resulting tensor light shift using 813\,nm and 521\,nm. The combined wavelengths yields a suppressed tensor light shift in \greenState.} 
    \label{fig:thelightshifts}
\end{figure}

Here, we consider the possibility of using the  large angular momentum  \greenStatefull $F=9/2$ state for fast gate operations by leveraging its large magnetic dipole compared to \clockStatefull~\cite{Omanakuttan_2021,siva_rydbgerg}. We define our computational basis (see Fig.~\ref{fig:overview}b): \groundStatefull $(g)$ spanning nuclear spin states $\ket{0}_g = \ket{g, m_I =-9/2},\cdots, \ket{9}_g = \ket{g, m_I =9/2}$; in \greenStatefull $F=9/2$ ($e$) the computational basis comprises hyperfine states $\ket{0}_e = \ket{e, m_F =-9/2},\cdots, \ket{9}_e =\ket{e, m_F =9/2}$. The ratio of $g$-factors between these two states is $\simeq 1.1 \times 10^{3}$, enabling fast rf-based control of qudits stored in $e$. The gate speed-up in \greenState comes at a price: tensor light shifts in $e$ impact coherence and limit gate times; see Fig.~\ref{fig:thelightshifts} for the tensor light shift scale induced by monochromatic light fields. Qudit-based quantum computing applications in \greenState require tensor light shift control to fully leverage qudits as competitive quantum computing resources. Finally, we want to underscore that while we choose \greenStatefull $F=9/2$ as the computational basis to match qudit size with \groundStatefull, we will show this technique enables magic trapping in all hyperfine levels in \greenStatefull \emph{simultaneously}.

To illustrate tweezer-induced dephasing, let us introduce the Hamiltonian defining this system as $H = H_{\text{Zeeman}} + H_{\text{LS}}$, where the Zeeman Hamiltonian is expressed as $H_{\text{Zeeman}} =  g_F \mu_B \,\vec{\mathbf{B}} \cdot \hat{\mathbf{F}}$;  $\vec{\mathbf{B}}$ is the magnetic field vector, $g_F$ is the Land\'e {\textit{g}-}factor, $\mu_B$ is the Bohr magneton and $\hat{\mathbf{F}}$ is the total angular momentum operator. Furthermore, in this paper we will make the distinction between $H_{\text{Zeeman}}$, which induces higher-order shifts via hyperfine mixing~\cite{boyd_nuclear_spins},  and $H^{(1)}_{\text{Zeeman}}$, which encodes the first-order term of the Zeeman Hamiltonian representing the linear Zeeman shift in the $m_F$ basis.

Now, we express the light shift Hamiltonian in its coordinate-invariant form~\cite{arnospaper} as,
\begin{widetext}
\begin{equation}
H_{\text{LS}} = -\frac{1}{4} \abs{E_0}^2 \Bigg(
    \, \alpha^s_{i}(\omega)
    + \alpha^v_{i}(\omega) \frac{i \left( \boldsymbol{\epsilon} \times \boldsymbol{\epsilon}^{*} \right) \cdot \hat{\mathbf{F}}}{2F} \\
    + \alpha^t_{i}(\omega) \frac{
        3 \left[
            \left( \boldsymbol{\epsilon}^{*} \cdot \hat{\mathbf{F}} \right) \left( \boldsymbol{\epsilon} \cdot \hat{\mathbf{F}} \right)
            + \left( \boldsymbol{\epsilon} \cdot \hat{\mathbf{F}} \right) \left( \boldsymbol{\epsilon}^{*} \cdot \hat{\mathbf{F}} \right)
        \right]
        - 2 \hat{\mathbf{F}}^2
    }{2F(2F-1)}
\Bigg)
\label{eq:invariant_ls_hamiltonian}
\end{equation}
\end{widetext}

\noindent where $\omega$ represents the angular frequency of the electric field,  $\mathbf{\epsilon}$ the polarization vector, $\alpha^s_{i}$, $\alpha^v_{i}$ and $\alpha^t_{i}$ correspond to the bare scalar, vector, and tensor polarizabilities (see Appendix~\ref{sec:appendix_hyperfine} for Eq.~\eqref{eq:bare_pols}), $E_0$ the peak value of the electric field, $ E_0(r=0) = \sqrt{\frac{4 P}{\pi c \epsilon_0 w_0^2}}$, $w_0$ is the $1/e^2$ beam radius of a Gaussian beam, $P$ is the optical power, $c$ is the speed of light in vacuum and $\mathbf{\epsilon}_0$ is the vacuum permittivity.

With linearly-polarized trapping light, the $\alpha^t_{i}$ term in Eq.~\ref{eq:invariant_ls_hamiltonian} dominates qudit dephasing, by creating unwanted qudit operations, impacting state fidelity. Let us express the ideal polarization vector as
\begin{equation}
\boldsymbol{\epsilon}(\beta) = [ \sin{\beta} ,\, 0 ,\,  \cos{\beta}].
\label{eq:polarization}
\end{equation}
 In the large magnetic field limit we can neglect off-diagonal terms in this Hamiltonian of Eqn.~\ref{eq:invariant_ls_hamiltonian}, which enables magic-angle tuning, where $\beta$ is set to suppress differential light shifts in a variety of settings~\cite{caltech_tweezers, katori_3p1, adam_3p1_bosonic, adam_3p1_bosonic_813, Yamamoto_magic_yb, 1064_magic_pol,monika_yb_magic_angle_höhn2024determining3p0excitedstatetuneout,joanna_omg}. This technique enables magic trapping conditions~\cite{decoupling_degrees_of_freedom, tobythesis} in wavelengths which traditionally would not support differential light shift cancellation. For qudits, this technique provides a path for removing tensor light shifts by setting $\beta$ to the tensor magic angle~\cite{tensor_magic_angle}, $\beta_0$ $\equiv$ 54.7$^\circ$. For quantum computing applications operating in \greenState the use of this angle for mitigating tensor light shifts requires (1) an impractical magnetic field magnitude (orders of $10^{3}$ G) to define the quantization axis, (2) an unachievable angular precision $\delta \beta = 2 \times 10^{-5}$ radians. Furthermore, magic-angle tuning is linearly sensitive to angular fluctuations, which is compounded by the scale of light shifts present in \greenState (see Fig.~\ref{fig:thelightshifts}), making operation of this technique on traditional tweezers unsuitable for qudit-based quantum computing in \greenState. 

In this paper we present an approach for harnessing tensor light shifts present in \greenState for engineering suitable magic trapping conditions for qudits via \textit{bichromatic} tweezers~\cite{ion_bic,2color_ions,2color_Lekien_waveguide,hilton_bichromatic,beloy_twocolor,2color_dimple,2color_odt,2color_odp_cornish,2color_quantum_memory,2color_saffman_cs,2frequency_compensation,arb_dipole_pol,anas_paper_scattering}. We consider two bichromatic tweezer configurations: (1) a proof-of-concept configuration of two scalar magic wavelengths; (2) a combination of the \groundState -- \clockState magic wavelength at 813.5\,\unit{\nano\meter}~\cite{jun_magic_813},  and a complementary wavelength near 521.3\,nm to yield magic trapping conditions for qudits in \groundState and \greenState, a 18-state manifold. Instead of relying solely on magic-angle tuning to mitigate differential light shifts, we use two distinct colors, as depicted in Fig.~\ref{fig:overview}c, at an appropriate power ratio, that together reduce total differential light shift in three ways: (1) choosing two wavelengths with opposite tensor light shift sign reduces the scale of light shifts significantly (see Fig.~\ref{fig:thelightshifts}), (2) balancing intensity to eliminate the net tensor shift, and (3) finally, tuning $\beta$ to $\beta_0$ with $\delta \beta = 2 \times 10^{-2}$ radians ($\delta \beta = 1^{\circ}$) and practical magnetic field magnitudes. Simultaneously implementing these conditions results in a highly robust tensor shift null in which off-diagonal elements of the light shift Hamiltonian are negligible, yielding an optical tweezer that is insensitive to power and angle fluctuations to second-order in these parameters.

\section{Engineering Bichromatic Optical Tweezer}
\label{sec:bic_formalism}

Our scheme tackles two challenges: (1) coherent transfer for qudits from $g$ to $e$ (see Fig.~\ref{fig:overview}b) and (2) elimination of tweezer-induced qudit operations in $e$ (see Fig.~\ref{fig:thelightshifts}). Both concerns require light shift engineering: the former requires achieving scalar and tensor magic trapping conditions while the latter requires the elimination of tensor light shifts to preserve qudit coherence. Our technique tackles magic trapping conditions for all magnetic sublevels in $e$ while also suppressing tensor-induced dephasing at a level of 99.9\% state fidelity. 

In Fig.~\ref{fig:thelightshifts} we present the incompatibility between light shifts in $g$, which is in the order of Hz (due to $J=0$ suppressing tensor light shifts) and in $e$. This mismatch means that a qudit prepared in $g$ and then coherently excited to $e$ (see Fig~\ref{fig:overview}b) will experience tweezer-induced dephasing from the tensor light shift. To mitigate this effect we need the following conditions: two distinct wavelengths with (1) opposite tensor polarizability sign (see Fig.~\ref{fig:negative_tensor} in Appendix~\ref{ssec:tensor_negative_sign}), and (2) an appropriate power ratio between the two colors such that resulting differential scalar and tensor light shifts can be suppressed, yielding scalar and tensor magic trapping conditions. Realizing scalar magic trapping enables high-fidelity coherent excitation~\cite{klusener} to \greenState (see Fig.~\ref{fig:overview}b), while tensor magic trapping (see Fig.~\ref{fig:thelightshifts}) preserves qudit coherence. 

\subsection{Effective Atomic Polarizabilities in Bichromatic Tweezers}
\label{ssec:bic_formalism}
Consider two co-propagating tweezers of two distinct wavelengths $\lambda_1$, $\lambda_2$ (see Fig.~\ref{fig:overview}c) with equal beam waist $w_0$, at optical powers $P_1$, $P_2$. Furthermore, we can parametrize $P_1, P_2$ via an offset $x$ between them scaled by a nominal power $P$~\cite{large_nuclear_spin_france,beloy_twocolor},

\begin{equation}
    P_{1,2} = (1 \pm x) P
\label{eq:mikestrick}
\end{equation}

Let us define the scalar magic condition,
\begin{equation}
    |\alpha_g^s(\lambda) - \alpha_e^s(\lambda)| = 0
\end{equation}

\noindent where $\alpha_i^s(\lambda)$ represents the bare scalar polarizability; see Eq.~\eqref{eq:bare_pols} in Appendix~\ref{sec:appendix_hyperfine}. For an electronic state $i$ (in this context, for $g$, $e$), $\lambda$ is the target wavelength that yields the minimal differential scalar polarizability. These conditions allow us to frame the search for viable parameters for the bichromatic tweezer as a light shift minimization problem. We can define the total scalar light shift in an electronic state as
\begin{equation}
    \bar \alpha_{i}^{s}(\lambda_1, \lambda_2, x) = (1+x) \alpha_i^s(\lambda_1) + (1-x) \alpha_i^s(\lambda_2)
    \label{eq:eff_scalar_pol}
\end{equation}

\noindent and total tensor shift
\begin{equation}
    \bar \alpha_{i}^{t}(\lambda_1, \lambda_2, x) =  (1+x) \alpha_i^t(\lambda_1) + (1-x) \alpha_i^t(\lambda_2), 
    \label{eq:eff_tensor_pol}
\end{equation}

\noindent where $\alpha_i^t(\lambda)$ represents bare tensor polarizability; see Eq.~\eqref{eq:bare_pols} in Appendix~\ref{sec:appendix_hyperfine}. 

Individually these conditions provide the framework to engineer either scalar magic or tensor magic potentials. For our application, we require both scalar and tensor magic,
\begin{equation}
   \bar \alpha_{e}^{s}(\lambda_1, \lambda_2, x_0)  = \bar \alpha_{g}^{s}(\lambda_1, \lambda_2, x_0) 
    \label{eq:min_bic_tweezer_scalar}
\end{equation}
\noindent and
\begin{equation}
   \bar \alpha_{e}^{t}(\lambda_1, \lambda_2, x_0)  = \bar \alpha_{g}^{t}(\lambda_1, \lambda_2, x_0) 
    \label{eq:min_bic_tweezer_tensor}
\end{equation}

\noindent In Eq.~\eqref{eq:min_bic_tweezer_tensor} we set the total tensor shift to zero given the scale of tensor polarizability in \groundState is on the order of $10^{-5}$ atomic units (a.u.)~\cite{shi_paper}. The resulting configuration $\lambda_1, \lambda_2, x_0$ yields an optical potential that is scalar and tensor magic for all $m_F$ in a hyperfine state. We define $x_0$ as the magic power ratio between the two wavelengths to engineer magic trapping conditions.

This framework illustrates the primary feature that a bichromatic potential provides for light shift engineering: the decoupling of the internal degrees of freedom of an atom via tuning parameters ($x, \lambda_1, \lambda_2$). These parameters enable $\bar \alpha_{i}^{t}(\lambda_1, \lambda_2, x) \approx 0$ (realizing the limit where $ H_{\text{Zeeman}}$ dominates), in the perturbative regime for a wide range of magnetic fields. Our technique's approach to realizing the perturbative regime is predicated on  scaling \textit{down} the tensor component of the light shift Hamiltonian (making it effectively zero) rather than on scaling \textit{up} the magnetic field amplitude (necessary for magic-angle tuning).

We now consider the angle sensitivity of magic-angle tuning in the perturbative regime. Light shift suppression relying solely on $\beta$ sets angle precision as the bottleneck for robust magic-angle tuning. We can describe the light shift sensitivity due to angle fluctuations,
\begin{equation}
\Bigg | \frac{ \partial E_{\text{LS}} } {\partial \beta} \Bigg |  \propto  I_0  \sin(2 \beta) \times \tilde{\alpha}_{i}^{t}
\label{eq:ls_sensitivity_prop}
\end{equation}
\noindent where $I_0$ is the total tweezer beam intensity; for a monochromatic tweezer $\tilde{\alpha}_{i}^{t} = \alpha_{i}^{t}(\lambda)$ and for a bichromatic tweezer $\tilde{\alpha}_{i}^{t} = \bar{\alpha}_{i}^{t}(\lambda_1, \lambda_2, x)$. Using Eq.~\ref{eq:ls_sensitivity_prop} we determine that light shift cancellation via bichromatic tweezers suppresses the total tensor light shift, where the overall scaling with $\bar{\alpha}_{i}^{t}$ is apparent. Thus the bichromatic approach vastly reduces the effect of angle fluctuations and enables the operation of the tensor magic angle at practical experimental angular precision.

In the following sections of this paper we assert the following experimental conditions used for results presented in this paper: (1) beam waists are equal for both wavelengths with a beam radius of 1 micron, and (2) the time scale of 1 millisecond is used for all state fidelity calculations. Finally, we note that in this paper we calculate the magic wavelength~\cite{jun_magic_813} for \clockState to be 813.46504\,\unit{\nm} using available reduced dipole matrix elements in the literature~\cite{caltech_tweezers,trautmann_quadrupole_transition,jun_rdme}. Lastly, we allow for systematic offset on wavelengths in the vicinity of 500\,nm of 100\,\unit{\mega \hertz}.
\begin{figure}[!h]
\centering
    \includegraphics{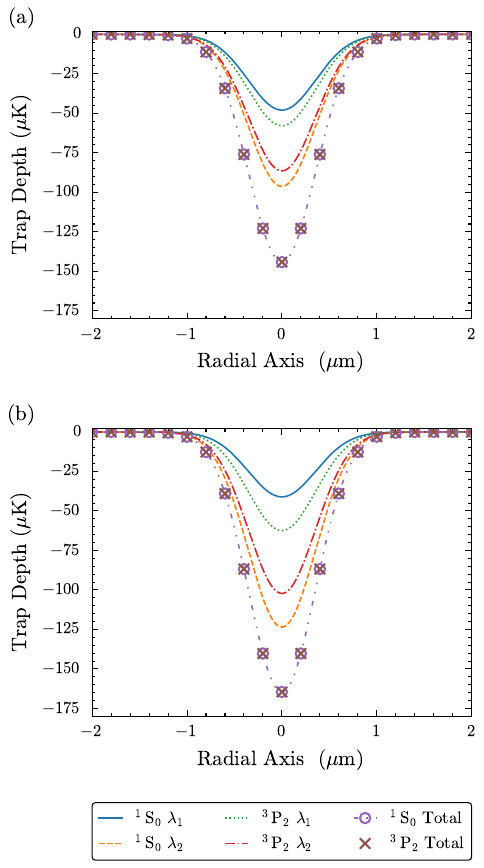}
        \caption{ \textbf{Bichromatic Tweezer Configurations in \Sr \greenState.} (a) Bichromatic tweezer using 891.5\,nm and 518.0\,nm for \groundState -- \greenState. In this figure $\lambda_1 = 891.5$\,nm and $\lambda_2 = 518.0$\,nm. The solid blue and the dotted green lines represent the optical potential contributions at 891.5\,nm for \groundState and \greenState respectively. The red dash-dot and the orange dashed lines represent the contributions for 518.0\,nm for \greenState and \groundState respectively. The total potential for \groundState is presented by the purple dashed line with unfilled circles and for \greenState is presented by brown cross markers.
        (b) Bichromatic tweezer using 813.5\,nm and 521.3\,nm for \groundState -- \greenState. In this figure $\lambda_1 = 813.5$\,nm and $\lambda_2 = 521.3$\,nm. The solid blue and the dotted green lines represent the optical potential contributions at 813.5\,nm for \groundState and \greenState respectively. The red dash-dot and the orange dashed lines represent the contributions for 521.3\,nm for \greenState and \groundState respectively. The total potential for \groundState is presented by the  purple line with unfilled circles and for \greenState is presented by brown cross markers.} 
    \label{fig:thetweezers}
\end{figure}

\subsection{Bichromatic Tweezers in \greenState}
\label{ssec:scalar_magic}
In this section we present two configurations for engineering bichromatic tweezers in \greenState: 891.5\,nm and 518.0\,nm and 813.5\,nm (the magic wavelength derived from our model) and 521.3\,nm. 

To illustrate the role of intensity compensation in removing tensor light shifts, we find two suitable scalar magic wavelengths (for $g$ and $e$) near 891.5\,nm and 518.0\,nm. Choosing scalar magic wavelengths reduces the task to a tensor light shift minimization problem using Eq.~\eqref{eq:min_bic_tweezer_tensor}, enabling the estimating of the appropriate power ratio to cancel out tensor light shifts. In Fig.~\ref{fig:thetweezers}a we present a bichromatic tweezer using 891.5\,nm and 518.0\,nm. In Table~\ref{tab:params_combined_tweezer} we present the optical powers to yield scalar and tensor magic for nuclear spins in both electronic states. We estimate a trap depth of 144\,\unit{\micro \kelvin}.

Now we turn our attention the engineering of a bichromatic tweezer in which one of its wavelengths is at 813.5\,nm. Qudit readout is a crucial component in our quantum computing scheme. The \clockState magic wavelength at 813.5\,nm~\cite{Ye_813} is a suitable wavelength for shelving atoms in \clockState under magic trapping conditions, while individual atoms are probed via the broad-line transition at 461\,nm. In all cases we consider, 813.5\,nm will be state readout and detection tweezer wavelength. In the 813.5\,nm/521.3\,nm bichromatic configuration, this readout stage is accomplished by the extinction of the 521.3\,nm tweezer. In the 891.5\,nm/518.0\,nm configuration this is accomplished through coherent transfer to an 813.5\,nm tweezer. In Table~\ref{tab:params_combined_tweezer} the optical power parameters in a bichromatic tweezer using 813.5\,nm and 521.3\,nm (see Fig.~\ref{fig:thetweezers}b). We calculate a trap depth of 164.56\,\unit{\micro\kelvin} and radial and axial trapping frequencies $\omega_r$, $\omega_a$ of 53.33\,\unit{\kilo\hertz} and 5.09\,\unit{\kilo\hertz} respectively.

\subsection{Experimental Tolerances Requirements for Qudit Manifold Fidelity in Bichromatic Tweezers}
\label{sec:hs_argument}
The challenge of using a monochromatic tweezer in \greenState lies in the scale of the tensor light shifts. Our technique enables us to leverage both the light shift cancellation achieved by using two wavelengths, which scales down the effective tensor polarizability, at a specific power ratio, which enables the operation at the tensor magic angle at practical magnetic fields. In this section we will estimate the necessary power offset tolerances to maintain a state fidelity of 0.999 at an angular precision of $1^{\circ}$. In what follows, we find that frequency accuracy of the tweezer beams at the level of 100\,\unit{\mega \hertz} does not impact yielding our target state fidelity.
\begin{figure}[!h]
\centering
    \includegraphics[width=1\columnwidth]{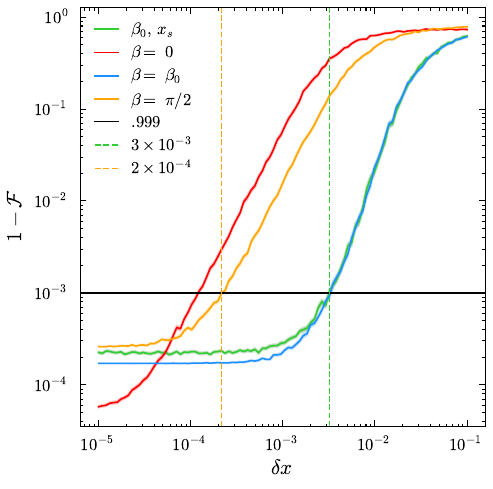}
        \caption{\textbf{State Infidelity as a Function of Power Ratio Precision for Bichromatic Tweezer using 891.5\,nm and 518.0\,nm.} We estimate the average state fidelity via a Monte Carlo simulation over 1000 trials per base point (100 points spanning from $\delta x = 10^{-5} $ to $10^{-1}$). The blue line represent estimated state fidelity at $(x_0, \beta_0)$. The orange line represents estimated state fidelity at $(x=x_0, \pi/2)$. The green line represents the state fidelity at $(x=x_0 + x_s, \beta_0)$. The red line represents the state fidelity at $(x=x_0, \, \beta_0)$. The green vertical line represents value for $\delta x$ to yield $\mathcal{F}=0.999$ for $(x=x_0 + x_s, \beta_0)$. The orange vertical dash-dotted line represents $\delta x$ to yield $\mathcal{F}=0.999$ for $(x=x_0 \, , \pi/2)$. Finally, the black horizontal line represents target fidelity $\mathcal{F}=0.999$. The band around each line represents the standard error around the mean for each configuration. The asymptotic tail at small $\delta x$ reflects the impact of the quadratic Zeeman shift.
}
    \label{fig:hs_overlap_comparison}
\end{figure}

We can estimate state fidelity via the Hilbert-Schmidt distance~\cite{tyler_hs, klaus_math},

\begin{equation}
\mathcal{F} =
\frac{1}{n_{\text{rel}}(n_{\text{rel}}+1)}\big[\text{Tr}(M_{\text{rel}}M_{\text{rel}}^\dagger)+|\text{Tr}(M_{\text{rel}})|^2\big]
\label{eq:fidelity_overlap_test}
\end{equation}

\noindent where $M_{\text{rel}}=PU_0^\dagger UP$, $P$ is the projection operator representing the computational basis in \greenState $F=9/2$, $U_0^\dagger U$ represents the overlap between the target unitary and realized unitary, of the form $ U = e^{-i t H}$, $n_{\text{rel}}$ represents the dimension of the computational basis ($n_{\text{rel}}=10$).For this calculation we are interested in the overlap between the target Hamiltonian, the linear Zeeman Hamiltonian $H^{(1)}_{\text{Zeeman}}$, and the realized Hamiltonian, defined as $ H_{\text{Realized}} = H_{\text{LS}} +H_{\text{Zeeman}}$, where the Zeeman Hamiltonian induces higher-order shifts via hyperfine mixing (such as the quadratic Zeeman shift~\cite{boyd_nuclear_spins}).
\begin{table}[!htbp]
\begin{tabular}{cccc}
\toprule
$x$ & $\lambda$ & $P_i$ & $\delta P_i / P_i$ \\
\midrule
$0.2573$
  & $\lambda_1 = 891.5\,\mathrm{nm}$ & $1.257\,\mathrm{mW}$ & 
  $2.3 \times 10^{-3}$ \\
  & $\lambda_2 = 518.0\,\mathrm{nm}$ & $0.743\,\mathrm{mW}$ & $3.9 \times 10^{-3}$ \\
\midrule
\addlinespace
\multirow{2}{*}{$7.467 \times 10^{-4}$}
  & $\lambda_1 = 813.5\,\mathrm{nm}$ & $1.001\,\mathrm{mW}$ & $2.4 \times 10^{-3}$ \\
  & $\lambda_2 = 521.3\,\mathrm{nm}$ & $0.999\,\mathrm{mW}$ & $2.4 \times 10^{-3}$ \\
\bottomrule
\label{tab:params_combined_tweezer}
\end{tabular}
\caption{Power Stability for Bichromatic Tweezer in \groundState--\greenState. The fractional power uncertainty for each beam is calculated by estimating the power ratio resolution $\delta x$. We define $\delta P_i / P_i = \delta x / (1 \pm x)$, where $P_i = (1\pm x) P$ represents the optical power of each color weighted by $P= 1\,\mathrm{mW}$. These values only assume $\delta \beta = 1^{\circ}$.}
\end{table}

In Fig.~\ref{fig:hs_overlap_comparison} we present the power ratio precision needed for yielding a bichromatic tweezer using two suitable scalar magic wavelengths for $\mathrm{B}=100$\,mG. This figure is obtained by tuning the power ratio uncertainty $\delta x$ over a range of 100 points spanning from $10^{-5}$ to $10^{-1}$. We assume uncertainties $\delta x , \delta \beta $ to be gaussian (these values represent the standard deviation used to sample a random error around $x, \beta$) used to compute state fidelity. Furthermore we considered a systematic offset error in the power ratio $x_s = 1\times 10^{-3}$ to estimate the impact on imperfections in relative power stabilization of the tweezer beams (such that the value used for estimating the impact of gaussian error is $x= x_0 + x_s$) as well as a systematic frequency offset $f_s = 100$MHz for 518.0\,nm to account for imperfection on laser stabilization (e.g. wavemeter calibration offsets). Moreover, we note that all configurations in the plot assume $\delta \beta = 1^{\circ}$. For each point $\delta x$ in this range, for a fixed $\delta \beta$, we estimate fidelity out of an average of 1000 trials. For this configuration we calculate the required fractional power uncertainties (at $\beta_0)$ to be $2 \times 10^{-3}$ and $4 \times 10^{-3}$ for 891.5\,nm and 518.0\,nm respectively; {for the configuration using 813.5\,nm and 521.3\,nm we find that by operating at $\beta_0$ a state fidelity of 0.999 can be realized at a fractional power uncertainty of $2 \times 10^{-3}$.} We note that operating at $\beta \neq \beta_0$ (such as $\beta=\pi/2$) increases the demands in power stabilization due to the increased loss (see vertical orange dashed line in Fig.~\ref{fig:hs_overlap_comparison}) of the compounded cancellation by operating \emph{simultaneously} at $x_0, \beta_0$.

\begin{figure}[!h]
\centering
    \includegraphics[width=1\columnwidth]{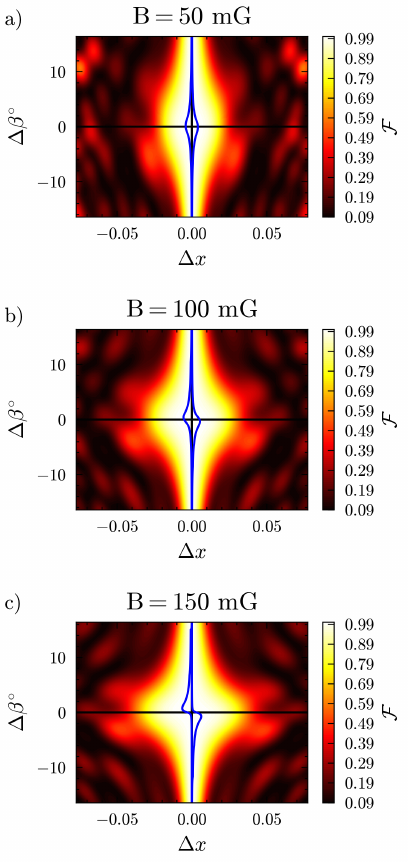}
        \caption{\textbf{State Fidelity as Function of Power Ratio and Quantization Axis Offset from Ideal Magic Conditions for 1 ms.} State fidelity (a) at $\mathrm{B}$ = $50$\,mG, (b) $\mathrm{B}$ = $100$\,mG, (c) $\mathrm{B}$ = $150$\,mG. The horizontal line represents the tensor magic angle while the vertical line represents the optimal power ratio. The region in the center represents the fidelity plateau. The blue contour represents the target fidelity of 0.999. All points within the blue contour exceed the target fidelity of 0.999.}
    \label{fig:hs_overlap_flat_3d}
\end{figure}


Armed with Eq.~\eqref{eq:fidelity_overlap_test} we can map the parameter space to estimate the response of fidelity as a function of offset $\Delta x, \Delta \beta$ from the optimal parameters $x_0, \, \beta_0$ for a fixed magnetic field. In Fig.~\ref{fig:hs_overlap_flat_3d} we present three different magnetic fields and its impact on fidelity for the configuration comprised of 891.5\,nm and 518.0\,nm. These magnetic fields choices are informed by the scale of the quadratic Zeeman shift~\cite{boyd_nuclear_spins} for the computational basis, which we calculate to be $\Delta^{2}_\mathrm{B} \approx (188.7\,\mathrm{Hz/G^2}) m_F^2 \mathrm{B}^2$ (see Appendix.~\ref{appendix:quadratic_zeeman}). At 50\,mG (Fig.~\ref{fig:hs_overlap_flat_3d}a), the fidelity yielded at $x \neq x_0$ is suppressed. In this regime the quantization axis is poorly defined. Increasing the magnetic field to 100\,mG (Fig.~\ref{fig:hs_overlap_flat_3d}b) shows 1) the perturbative regime signified by the broadening around $\beta_0$ of nonzero fidelity and 2) the impact of the quadratic Zeeman shift as shown by the slight shearing of the contour for 0.999 qudit fidelity, an effect of the interplay between the Zeeman shift and the light shift~\cite{magneto_optic_shift}; see Eq.~\eqref{eq:second_order_shift_overlap} in Appendix~\ref{appendix:pt_notes}. This shift is introduced by the overlap between the light shift, when operating at $(x, \beta) \neq (x_0, \beta_0)$, and the Zeeman Hamiltonian, yielding a $m_F^3$ term, which we will refer to as a magento-optic shift. Moreover, at both edges of the figure (representing monochromatic potentials) fidelity remains suppressed. Finally, at 150\,mG (Fig.~\ref{fig:hs_overlap_flat_3d}c), the magnetic field induced dephasing leads to a reduction in the parameter space for yielding qudit fidelity of 0.999. We find that the shape of the plateau highlights the trade-off between power and angle stability. Given this result we expect that operating at $\mathrm{B}\approx$ 100\,mG affords favorable conditions for our scheme given the that fidelity peak can be accessed through a wider set of possible configurations (in all four quadrants of the parameter space) and angular and power stability as opposed to 157\,mG, the maximum magnetic field to maintain 0.999 fidelity.

Our analysis highlights the light shift engineering realized by bichromatic tweezers: it realizes a light shift regime in which the off-diagonal elements in Eq.~\eqref{eq:invariant_ls_hamiltonian} are small compared to the diagonal elements. This is the perturbative regime in which magic-angle tuning \emph{can be engineered}, enabling the preservation of qudit coherence by suppressing dephasing induced by the trapping potential. Furthermore, the light shift engineering achieves 1) mitigation of higher-order shifts introduced by hyperfine mixing, 2) the decoupling of the magnetic and light-matter interaction via the suppression magneto-optic shift. Finally, we estimate the magnetic field level in which the remaining dephasing mechanism in this system, quadratic Zeeman shift, can be mitigated to maintain our target qudit fidelity while operating at achievable angular precision.
\section{Tweezer-Induced Decoherence in Bichromatic Tweezers}

\begin{figure}[!h]
\centering
\includegraphics{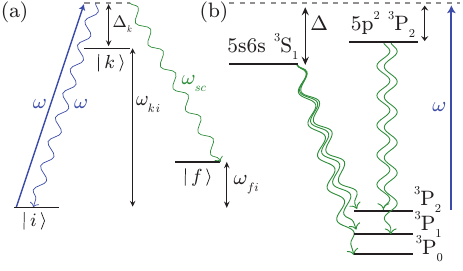}
\caption{\textbf{Tweezer-Induced Scattering Processes in \greenState} (a) Optical tweezers' far-detuned light is represented by the blue line (corresponding for $\omega$) connecting a trapped qudit in the computational state $\ket{i}$ with dipole-allowed transitions to excited state $\ket{k}$, which introduce virtual states, detuned by $\Delta_k$ from $\ket{k}$ indicated by dashed gray lines. These states open decay (error) channels for qudits. When scattered photons (curly line) maintain frequency $\omega$, this represents a Rayleigh scattering event. In contrast, a Raman scattering occurs when scattered photons change color (from blue to green) $\omega \rightarrow \omega_{\text{sc}}$. This process enables a qudit to decay to a different state  $\ket{f}$. (b) Leakages Errors in \greenState. In \greenState, excited states presented in this figure create decay paths to other metastable states in \Sr such as \clockState and \redState that a qudit can access by Raman scattering.}
\label{fig:raman_rayleigh_diagram}
\end{figure}

Optical tweezers (at frequency $\omega$, see Fig.~\ref{fig:raman_rayleigh_diagram}) induce via Rayleigh and Raman scattering that impact qudit coherence. Rayleigh scattering (see Fig.~\ref{fig:raman_rayleigh_diagram}a) is an elastic process~\cite{lisdat_scattering,mike_thesis} that does not change the atom's internal state and is characterized by scattered photons' maintaining frequency $\omega$. However, for \greenState, this process yields differential scattering amplitudes between nuclear spins in \Sr, introducing a dephasing mechanism~\cite{uys_elastic_rayleigh,era_joanna}. In contrast, Raman scattering is an inelastic process (see Fig.~\ref{fig:raman_rayleigh_diagram}a) characterized by a change of \textit{color}, $\omega \rightarrow \omega_{sc}$, which changes the atom's internal state because of spontaneous optical pumping. This process leads to the emergence of depolarizing errors and leakage errors~\cite{caltech_erasure} (see Fig.~\ref{fig:raman_rayleigh_diagram}b).

\subsection{Leakage Errors}
In this section we present the scattering rates for bichromatic tweezers. For both configurations presented in this work (813.5\,nm - 521.3\,nm and 891.5\,nm - 518.0\,nm) we set $\beta$ to $\beta_0$; the respective trap depth are 164\,\unit{\micro \kelvin} and 144\,\unit{\micro \kelvin} respective; the $1/e^2$ beam waist radius is 1 micron for both configurations. Finally, we note that a tweezer using 891.5\,nm and 518.0\,nm, yields an effective 327\,nm photon, below the ionization threshold. 

In Table~\ref{tab:scattering_rate_combined_813_521} and Table~\ref{tab:scattering_rate_combined_891_518} we present the leakage errors within \greenState for the two configuratins presented in this work. We calculate the scattering rates for bichromatic tweezers as follows: (1) we compute the individual scattering rates for each wavelength; (2) combine the rates to yield the total scattering rate; (3) take the average to yield the scattering rate process for a qudit in \Sr. We note that for the case $F'=9/2$ we look for depolarization channels in which the state undergoes $\ket{F,m_F} \rightarrow \ket{F, m_{F'}}$ where $m_F \neq m_F'$. 

\begin{table}[!htbp]
\begin{tabular}{cc}
\toprule
   $\ket{^3\mathrm{P}_2 , F=9/2} \rightarrow  \ket{^3\mathrm{P}_2 , F'}$ & $\Gamma$\\
\midrule
 $F'=5/2$ & 0.020 \\
 $F'=7/2$ & 0.025 \\
 $F'=9/2$  & 0.018  \\
 $F'=11/2$ & 0.044\\
 $F'=13/2$ & 0.015  \\
\midrule
   $\ket{^3\mathrm{P}_2 , F=9/2} \rightarrow \ket{^3\mathrm{P}_\mathrm{J} , F'}$  &   $\Gamma$ \\
\midrule
$^3\mathrm{P}_1$\\
 $F'=7/2$ & 0.079\\
 $F'=9/2$  & 0.137 \\
 $F'=11/2$ & 0.049\\
 \\
$^3\mathrm{P}_0$\\
 $F'=9/2$  & 0.066\\
\bottomrule
\end{tabular}
\caption{Average Raman Scattering Rates within $\mathrm{^3P_2}$ and to other metastable states $\mathrm{^3P_J}$ in Bichromatic Tweezer at $\beta_0$ using 813.5\,nm and 521.3\,nm. $\Gamma$ is in units of photons/sec.}
\label{tab:scattering_rate_combined_813_521}
\end{table}

\begin{table}[!htbp]
\centering
\begin{tabular}{cc}
\toprule
   $\ket{^3\mathrm{P}_2 , F=9/2} \rightarrow  \ket{^3\mathrm{P}_2 , F'}$ & $\Gamma$\\
\midrule
 $F'=5/2$ & 0.014\\
 $F'=7/2$ & 0.018\\
 $F'=9/2$  & 0.012  \\
 $F'=11/2$ & 0.031\\
 $F'=13/2$ & 0.011  \\
\midrule
   $\ket{^3\mathrm{P}_2 , F=9/2} \rightarrow \ket{^3\mathrm{P}_\mathrm{J} , F'}$  &   $\Gamma$ \\
\midrule
$^3\mathrm{P}_1$\\
 $F'=7/2$ & 0.068\\
 $F'=9/2$  & 0.133 \\
 $F'=11/2$ & 0.039\\
 \\
$^3\mathrm{P}_0$\\
 $F'=9/2$  & 0.057\\
\bottomrule
\end{tabular}
\caption{Average Raman Scattering Rates within $\mathrm{^3P_2}$ and to other metastable states $\mathrm{^3P_J}$ in Bichromatic Tweezer at $\beta_0$ using 891.5\,nm and 518.0\,nm. $\Gamma$ is in units of photons/sec.}
\label{tab:scattering_rate_combined_891_518}
\end{table}

\subsubsection{Blackbody Radiation-induced Optical Pumping}  
The blackbody spectrum at room temperature introduces an additional error channel:  optical pumping induced by blackbody radiation (BBR)~\cite{lisdat_bbr,optical_pumping_bbr,hill_phd_thesis_sr,3p2_katori_bbr}. In \greenStatefull this enables leakage to $^{3}\mathrm{D}_{\mathrm{J}}\xspace$ manifold. We can represent this error rate as a function of temperature, 

\begin{equation}
     \mathlarger{\mathlarger{\Gamma}} _{\text{BBR}, i\rightarrow f}(T) =\sum_{k} b_{i,k} b_{k, f}
    \frac{A_{ik}}{e^{\hbar \omega_{i,k}/k_{B}T}-1},
    \label{eq:bbr_pumping}
\end{equation}   
In Eq.~\eqref{eq:bbr_pumping}, $A_{ik}$ represents the Einstein A-coefficient $i\rightarrow k$, $\omega_{i,k}$ is the angular frequency between the two states of interest, $i \equiv \{ S_i, L_i, J_i \}$ encodes the angular momentum degrees of freedom, and $b_{i,k}$ represent the branching ratios~\cite{mike_thesis,hill_phd_thesis_sr,ludlowthesis},
\begin{equation}
b_{i,k}= (2 J_i + 1)(2L_k +1) \, \bigg\{\begin{array}{ccc}
L_i & J_i & S_i \\
J_k & L_k&1 
\end{array}\bigg\}^2
\end{equation}

We are interested in estimating the leakage rate for $e\rightarrow k \rightarrow f$ error channel (where $k = 5s4d \ \mathrm{^{3}D_J}$ and $f=$\, \redState). Using corresponding branching ratios~\cite{hill_phd_thesis_sr} and the radiative decay rate~\cite{safronova_bbr,3p2_katori_bbr}, this process yields a quenching rate of $8.03 \times 10^{-3}$ photons/sec~\cite{3p2_katori_bbr} at $T=299.5$\,K. 

\subsection{Photoionization in \greenState at 521.3\,nm}

The use of wavelengths in the vicinity of 521\,nm comes with a drawback: the introduction a loss channel via two-photon photoionization in metastable states. The wavelengths required to photo-ionize in \Sr are 218\,nm, 317\,nm, 318\,nm, and 322\,nm for \groundState, \clockState, \redState and \greenState respectively. Two photons with a combined effective energy above the ionization threshold allow for an ionization out of \metastablemanifold to the continuum~\cite{aaron_thesis}. This effect has been observed in 520\,nm tweezers~\cite{will_metasurfaces,caltech_tweezers,zeiher_520nm_loss_3p0_ionization} for \clockState and \redState. For \greenState our bichromatic tweezers operating at 813.5\,nm and 521.3\,nm, wavelengths which yield an effective energy higher than the ionization threshold, introduce this loss mechanism.

For the purposes of benchmarking this process in our tweezer in \greenState, we will extrapolate from the ionization rate in \clockState measured in F. Gyger et.\,al.~\cite{zeiher_520nm_loss_3p0_ionization}, $250\,\unit{\sec}^{-1}\,\unit{\milli\kelvin}^{-2}$. Furthermore, we will use this rate for our 521.3\,nm tweezer. The trap depth contribution for 521.3\,nm is $123\,\unit{\micro\kelvin}$, yielding a photoionization rate of $3.8\,\unit{\sec}^{-1}$. There is a second photoionization process combining a 813.5\,nm photon and a 521.3\,nm photon that will increase the photoionization rate, which we have not carefully considered. If this process yields a rate that becomes the limiting factor for our approach, it can be ameliorated by operating at different set of wavelengths, such as 891.5\,nm and 518.0\,nm, where only the shorter wavelength photons can combine to exceed the ionization threshold. See Table.~\ref{tab:scattering_rate_combined_891_518} for the corresponding scattering rates for this configuration. Given our expected gate speeds, less than 1\,ms, we do not expect these processes to be limiting factors for our scheme. 

Other issues with tweezers in the vicinity of 521\,nm have been identified as opening a loss channel on the $^{1}\mathrm{P}_{1} \rightarrow \mathrm{^{1}D}_{2}$ decay that occurs when cycling \groundState$\rightarrow ^{1}\mathrm{P}_{1}$, essential for state detection in our scheme. This loss mechanism arises because $^{1}\mathrm{D}_{2}$ is anti trapped near the vicinity of 515\,nm; see Fig.~\ref{fig:decay_path_pol} in Appendix~\ref{sec:1d2_leaks}. In contrast, for our scheme at 521.3\,nm (and 518.0\,nm), $^{1}\mathrm{D}_{2}$ is 
trapped, mitigating this loss channel.

\subsection{Rayleigh Decoherence in \greenState}
Unlike \clockState, the presence of resolvable hyperfine structure in \greenState allows for differential scattering amplitudes between nuclear spins, leading to Rayleigh-induced decoherence~\cite{era_uys,moorethesis}. For qubits this effect is represented by a sole figure of merit: differential scattering amplitude between two nuclear spins comprising a qubit; e.g. a hyperfine qubit $\frac{1}{\sqrt{2}}(\ket{0}+\ket{1})$ in Yb~\cite{era_joanna}. In contrast, qudits' symmetry structure require a modification to the result in H. Uys et.\,al.~\cite{era_uys}. We define an effective Rayleigh scattering for each nuclear spin in the hyperfine manifold using Eq.~\eqref{eq:effective_eva} in Appendix~\ref{sec:scattering_theory}. We will calculate this figure of merit to estimate impact on qudit coherence. 

We present the results of this calculation in Table.~\ref{tab:merged_scattering_ray_rates_green} for the two configurations presented in this paper. The effective Rayleigh rates are computed as follows: (1) we calculate the given difference in Rayleigh scattering between a given $i \equiv \ket{F, m_F}$ hyperfine state in the qudit (e.g $m_F = 9/2$) and $j \equiv \ket{F, m_{F'}}$ nuclear spin in the qudit; (2) exhaustively go through all remaining $j$ nuclear spins in the manifold  (e.g $m_F' = 7/2 \, , \dots,\, -9/2$); (3) take the average. 

\begin{table}[!htbp]
\centering
\begin{tabular}{ccc}
\toprule
\multicolumn{3}{c}{\textbf{813.5\,nm and 521.3\,nm}} \\
\midrule
   $m_F$ & $\Gamma_{\text{Scattering}}$ & $\Gamma_{\text{Decoherence}}$ \\
\midrule
 $\pm 9/2$ & 3.906 & 0.064 \\
 $\pm 7/2$ & 3.871 & 0.029 \\
 $\pm 5/2$ & 3.864 & 0.023 \\
 $\pm 3/2$ & 3.869 & 0.028 \\
 $\pm 1/2$ & 3.875 & 0.033 \\
\addlinespace[1.5ex]
\midrule
\multicolumn{3}{c}{\textbf{891.5\,nm and 518.0\,nm}} \\
\midrule
   $m_F$ & $\Gamma_{\text{Scattering}}$ & $\Gamma_{\text{Decoherence}}$ \\
\midrule
 $\pm 9/2$ & 3.705 & 0.045 \\
 $\pm 7/2$ & 3.681 & 0.020 \\
 $\pm 5/2$ & 3.676 & 0.016 \\
 $\pm 3/2$ & 3.679 & 0.019 \\
 $\pm 1/2$ & 3.683 & 0.023 \\
\bottomrule
\end{tabular}
\caption{Comparison of traditional Rayleigh scattering and decoherence error rates in Bichromatic Tweezers for \greenState~$F=9/2$ at $\beta_0$. All $\Gamma$ values are in units of photons/sec.}
\label{tab:merged_scattering_ray_rates_green}
\end{table}

By operating at the magic angle $\beta_0$, the bichromatic approach enables the mitigation of Rayleigh decoherence. In Table.~\ref{tab:merged_scattering_ray_rates_green} we estimate both Rayleigh scattering and decoherence rates. At $\beta_0$, $d_{\pm 1}$ contributions need to be considered in the scattering calculation (see Appendix.~\ref{ssec:arbitrary_pi_light}) in addition to $d_{0}$ contributions.

\section{Conclusion}
We have shown the viability of engineering magic trapping conditions for all $m_F$ in \greenStatefull $F=9/2$ of \Sr via bichromatic tweezers by carefully choosing a power ratio between two wavelengths and magic-angle tuning to the tensor magic angle. While our focus is on \greenStatefull $F=9/2$, we find that this technique enables magic trapping for all hyperfine levels in \greenStatefull \emph{simultaneously}. We apply this technique to engineer magic trapping conditions in \groundState -- \greenState with two configurations: (1) employing two scalar magic wavelengths at 891.5\,nm and 518.0\,nm, and (2) using the \groundState -- \clockState magic wavelength at 813.5\,\unit{\nano \meter} and an additional wavelength at 521.3\,\unit{\nano \meter}. 

We determine that for a high-spin hyperfine manifold a monochromatic tweezer is incompatible with magic-angle tuning: this technique requires magnetic fields well within the Paschen-Back regime (where $\ket{F, m_F}$ is no longer a good quantum basis) to define the quantization axis. We have shown that operating at $\mathrm{B}\approx 100$\,mG would enable robust suppression of tweezer-induced dephasing with bichromatic tweezers while also mitigating the impact of the quadratic Zeeman shift~\cite{boyd_nuclear_spins} on qudit fidelity. Furthermore, since our light shift cancellation relies on scaling \emph{down} the light shift Hamiltonian rather than scaling \emph{up} the bias magnetic field, the perturbative regime needed for magic-angle tuning can be enforced at low magnetic fields. These conditions enable operating at the magic angle at practical angular precision, leading to the suppression of light shift induced-dephasing and Rayleigh decoherence --- key ingredients in preserving qudit coherence. We expect the bichromatic tweezer to enhance state fidelity, and enable the assembly of atomic arrays ideal for qudit-based quantum sensing, simulation, and computing applications in \Sr.

We thank M.\,Safronova for providing reduced dipole matrix elements for \groundState and \bluemotleak; K.\,Gan, B.\,Madhusudhana and S.\,Alperin for advising on considerations regarding the viability of this scheme throughout this project; B.\,Hunt, S.\,Pampel, Y.\, Liu, T\,.Hoang, and E\,.Gurra for meaningful conversations during the preparation of this manuscript. 

This work was supported by the Laboratory Directed Research and Development program of Los Alamos National Laboratory under project numbers 20210116DR, 20210955PRD3, and 20240295ER, the NSF Quantum Leap Challenge Institutes program, Award No. 2016244, and presented at the Aspen Center for Physics, which is supported by National Science Foundation grant PHY-2210452.

\bibliography{bib.bib}
\appendix
\section{Hyperfine Polarizability in \Sr}
\label{sec:appendix_hyperfine}
\renewcommand{\thefigure}{A\arabic{figure}}
\renewcommand{\thetable}{A\arabic{table}}
\setcounter{figure}{0}
\counterwithin{figure}{section}
\setcounter{table}{0}
\counterwithin{table}{section}
In this section we present the formalism used to estimate the atomic polarizabilities~\cite{LeKien2012DynamicalPO,steck,caltech_thesis} for \Sr used in this paper to find the appropriate wavelengths for light shift cancellation. In this paper computed the scalar, vector, and tensor bare hyperfine polarizabilities using following expressions,
\begin{widetext}
\begin{align}
    \alpha^s_{i}(\omega) &=  \frac{2}{3\hbar} \frac{1}{(2F+1)} \sum_{n'F'}  \langle n \, F'\|\mathbf{d}\|n \, F\rangle|^2
    \frac{\omega_{n'F'nF}}{\omega_{n'F'nF}^2 - \omega^2} \\[10pt]
    \alpha^v_{i}(\omega) &=  \frac{1}{\hbar} \sum_{n'F'} \sqrt{\frac{24F}{(F+1)(2F+1)}} (-1)^{F'+F+1} 
    \begin{Bmatrix}
        1 & 1 & 1 \\
        F & F' & F
    \end{Bmatrix}
    \langle n \, F'\|\mathbf{d}\|n \, F\rangle|^2
    \frac{\omega}{\omega_{n'F'nF}^2 - \omega^2} \\[10pt]
    \alpha^t_{i}(\omega) &=  \frac{1}{\hbar} \sqrt{\frac{40F (2F-1)}{3(F+1)(2F+1)(2F+3)}} 
    \sum_{n'F'} (-1)^{F'+F} 
    \begin{Bmatrix}
        1 & 2 & 1 \\
        F & F' & F
    \end{Bmatrix}
    \langle n \, F'\|\mathbf{d}\|n \, F\rangle|^2
    \frac{\omega_{n'F'nF}}{\omega_{n'F'nF}^2 - \omega^2}
    \label{eq:bare_pols}
\end{align}
\end{widetext}
\noindent $\omega_{n'F'nF}$ represents frequency detunings for each of the $k$ excited states for a given electronic state. Furthermore, $\langle n \, F'\|\mathbf{d}\|n \, F\rangle|$ represent hyperfine structure reduced matrix elements (RDMEs), which can be computed from known $\langle n \, J'\|\mathbf{d}\|n \, J\rangle|$ (see Tables~\labelcref{tab:rdme_ground,tab:rdme_red,tab:rdme_green}) via~\cite{budker,budker_book_ls,king2008angular},
\begin{widetext}
\begin{equation}
\langle n'\,J'\,I\,F'\,\|\mathbf{d}\|n\, J\,I\,F\,\rangle \notag \\
\quad = (-1)^{J'+I+F+1} \sqrt{(2F+1)(2F'+1)} 
\left\{
\begin{array}{ccc}
F' & 1 & F \\
J & I & J'
\end{array}
\right\} 
\\
\langle n J'\|\mathbf{d}\|n J\rangle
\end{equation}
\label{eq:rdme_f_arno}
\end{widetext}

\begin{table}[!ht]
    \centering
    \raggedbottom
    \allowdisplaybreaks
    \caption{Relevant Wavelength and RDMEs for the Excited Manifold for \groundState and \clockState.}
    \begin{tabular}{ccc}
        \toprule 
       State $k$ & $\Delta E_{ki}$ (\si{\per\centi\meter}) & $\Bra{k} D \Ket{i}$ (a.u.) \\
        \midrule
              &  State $i=$   $ ^{1}\mathrm{S}_{0}$ \\
              \\
        $5s5p \ ^{3}\mathrm{P}_{1}$ & 14504.3380 & 0.1508~\cite{safronovapriv} \\
        $5s6p \ ^{3}\mathrm{P}_{1}$ & 33868.0000 & 0.0340~\cite{safronovapriv} \\
        $5s5p \ ^{1}\mathrm{P}_{1}$ & 21698.4520 & 5.2479~\cite{safronovapriv} \\
        $5s6p \ ^{1}\mathrm{P}_{1}$ & 34098.4040 & 0.2664~\cite{safronovapriv} \\
        $5s7p \ ^{1}\mathrm{P}_{1}$ & 38906.8580 & 0.3650~\cite{safronovapriv} \\
        $5s8p \ ^{1}\mathrm{P}_{1}$ & 42462.1360 & 0.5900~\cite{jun_rdme} \\
        $5s9p \ ^{1}\mathrm{P}_{1}$ & 43328.0400 & 0.4575~\cite{jun_rdme} \\
        $5s10p \ ^{1}\mathrm{P}_{1}$ & 43938.2010 & 0.3394~\cite{jun_rdme} \\
        $5s11p \ ^{1}\mathrm{P}_{1}$ & 44366.4200 & 0.2505~\cite{jun_rdme} \\
        $5s12p \ ^{1}\mathrm{P}_{1}$ & 44675.7370 & 0.1996~\cite{jun_rdme} \\
        $5s13p \ ^{1}\mathrm{P}_{1}$ & 44903.5000 & 0.1602~\cite{jun_rdme} \\
        $5s14p \ ^{1}\mathrm{P}_{1}$ & 45075.2900 & 0.1375~\cite{jun_rdme} \\
        $5s15p \ ^{1}\mathrm{P}_{1}$ & 45207.8300 & 0.1167~\cite{jun_rdme} \\
        $4d5p \ ^{1}\mathrm{P}_{1}$ & 41172.0540 & 0.6005~\cite{jun_rdme} \\
        Rydberg $ \ ^{1}\mathrm{P}_{1}$ & 45932.2036 & 0.7037~\cite{jun_rdme} \\
        \\
             &  State $i=$ \clockState \\
              \\
        $5s6s \ ^{3}\mathrm{S}_{1}$ & 14721.2660 & 1.9718~\cite{jun_rdme} \\
        $5s7s \ ^{3}\mathrm{S}_{1}$ & 23107.1680 & 0.6099~\cite{jun_rdme} \\
        $5s8s \ ^{3}\mathrm{S}_{1}$ & 26443.8650 & 0.2735~\cite{jun_rdme} \\
        $5s9s \ ^{3}\mathrm{S}_{1}$ & 28133.6530 & 0.1849~\cite{jun_rdme} \\
        $5s10s \ ^{3}\mathrm{S}_{1}$ & 29109.9330 & 0.1373~\cite{jun_rdme} \\
        $5p^{2} \ ^{3}\mathrm{P}_{1}$ & 21082.5980 & 2.4824~\cite{jun_rdme} \\
        $4d^{2} \ ^{3}\mathrm{P}_{1}$ & 30278.4130 & 1.6216~\cite{jun_rdme} \\
        $5s4d \ ^{3}\mathrm{D}_{1}$ & 3841.5330 & 2.6906~\cite{jun_rdme} \\
        $5s5d \ ^{3}\mathrm{D}_{1}$ & 20689.4010 & 2.7249~\cite{jun_rdme} \\
        $5s6d \ ^{3}\mathrm{D}_{1}$ & 25368.3230 & 1.1388~\cite{jun_rdme} \\
        $5s7d \ ^{3}\mathrm{D}_{1}$ & 27546.8470 & 0.7537~\cite{jun_rdme} \\
        $5s8d \ ^{3}\mathrm{D}_{1}$ & 28749.1930 & 0.5475~\cite{jun_rdme} \\
        $5s9d \ ^{3}\mathrm{D}_{1}$ & 29487.3830 & 0.4238~\cite{jun_rdme} \\
        Rydberg $\ ^{3}\mathrm{S}_{1}$ & 31614.6966 & 0.2904~\cite{jun_rdme} \\
        Rydberg $\ ^{3}\mathrm{D}_{1}$ & 31614.6966 & 0.4247~\cite{jun_rdme} \\

        \bottomrule
    \end{tabular}
    \label{tab:rdme_ground}
\end{table}

\begin{table}[!ht]
    \centering
     \caption{Relevant Wavelength and RDMEs for the Excited Manifold for \redState. $\alpha_c, \alpha_o = 5.6, 42.2$ a.u. taken from Trautmann Thesis~\cite{trautmann_thesis}.}
    \begin{tabular}{ccc}
\toprule
        State $k$ & $\Delta E_{ki}$ (\si{\per\centi\meter}) & $\Bra{k} D \Ket{i}$ (a.u.) \\
\midrule
        $5s^{2} \ ^{1}\mathrm{S}_{0}$\xspace & -14505.000 & 0.151~\cite{caltech_tweezers} \\
        $5s4d \ ^{3}\mathrm{D}_{1}$\xspace & 3655.000 & 2.322~\cite{caltech_tweezers} \\
        $5s4d \ ^{3}\mathrm{D}_{2}$\xspace & 3714.000 & 4.019~\cite{caltech_tweezers} \\
        $5s4d \ ^{1}\mathrm{D}_{2}$\xspace & 5645.000 & 0.190~\cite{caltech_tweezers} \\
        $5s6s \ ^{3}\mathrm{S}_{1}$\xspace & 14534.000 & 3.425~\cite{caltech_tweezers} \\
        $5s6s \ ^{1}\mathrm{S}_{0}$\xspace & 16087.000 & 0.045~\cite{caltech_tweezers} \\
        $5s5d \ ^{1}\mathrm{D}_{2}$\xspace & 20223.000 & 0.061~\cite{caltech_tweezers} \\
        $5s5d \ ^{3}\mathrm{D}_{1}$\xspace & 20503.000 & 2.009~\cite{caltech_tweezers} \\
        $5s5d \ ^{3}\mathrm{D}_{2}$\xspace & 20518.000 & 3.673~\cite{caltech_tweezers} \\
        $5p^{2} \ ^{3}\mathrm{P}_{0}$\xspace & 20689.000 & 2.657~\cite{caltech_tweezers} \\
        $5p^{2} \ ^{3}\mathrm{P}_{1}$\xspace & 20896.000 & 2.362~\cite{caltech_tweezers} \\
        $5p^{2} \ ^{3}\mathrm{P}_{2}$\xspace & 21170.000 & 2.865~\cite{caltech_tweezers} \\
        $5p^{2} \ ^{1}\mathrm{D}_{2}$\xspace & 22457.000 & 0.228~\cite{caltech_tweezers} \\
        $5p^{2} \ ^{1}\mathrm{S}_{0}$\xspace & 22656.000 & 0.291~\cite{caltech_tweezers} \\
        $5s7s \ ^{3}\mathrm{S}_{1}$\xspace & 22920.000 & 0.921~\cite{caltech_tweezers} \\
        
        $5s7s \ ^{1}\mathrm{S}_{0}$\xspace & 23940.054 & 0.250~\cite{1064_magic_pol} \\
        $5s6d \ ^{3}\mathrm{D}_{1}$\xspace & 25181.305 & 0.986~\cite{1064_magic_pol} \\
        $5s6d \ ^{3}\mathrm{D}_{2}$\xspace & 25186.379 & 1.708~\cite{1064_magic_pol} \\
        $5s8s \ ^{3}\mathrm{S}_{1}$\xspace & 26257.057 & 0.479~\cite{1064_magic_pol} \\
        $5s7d \ ^{3}\mathrm{D}_{1}$\xspace & 27359.781 & 0.661~\cite{1064_magic_pol} \\
        $5s7d \ ^{3}\mathrm{D}_{2}$\xspace & 27365.022 & 1.145~\cite{1064_magic_pol} \\
        $5s9s \ ^{3}\mathrm{S}_{1}$\xspace & 27947.012 & 0.324~\cite{1064_magic_pol} \\
        $5s8d \ ^{3}\mathrm{D}_{1}$\xspace & 28562.452 & 0.480~\cite{1064_magic_pol} \\
        $5s8d \ ^{3}\mathrm{D}_{2}$\xspace & 28565.715 & 0.831~\cite{1064_magic_pol} \\
        $5s10s \ ^{3}\mathrm{S}_{1}$\xspace & 28923.469 & 0.240~\cite{1064_magic_pol} \\
        $5s9d \ ^{3}\mathrm{D}_{1}$\xspace & 29303.171 & 0.643~\cite{1064_magic_pol} \\
        $5s9d \ ^{3}\mathrm{D}_{2}$\xspace & 29303.171 & 0.371~\cite{1064_magic_pol} \\
        $4d^{2} \ ^{3}\mathrm{P}_{0}$\xspace & 30021.916 & 1.680~\cite{1064_magic_pol} \\
        $4d^{2} \ ^{3}\mathrm{P}_{1}$\xspace & 30092.384 & 1.720~\cite{1064_magic_pol} \\
        $4d^{2} \ ^{3}\mathrm{P}_{2}$\xspace & 30226.091 & 2.210~\cite{1064_magic_pol} \\

\bottomrule
    \end{tabular}
    \label{tab:rdme_red}
\end{table}

\begin{table}[!ht]
    \centering
   \caption{Relevant Wavelength and RDMEs for the Excited Manifold for \greenState of Sr.}
    \begin{tabular}{ccc}
\toprule
        State $k$ & $\Delta E_{ki}$ (\si{\per\centi\meter}) & $\Bra{k} D \Ket{i}$ (a.u.) \\
\midrule
        $5s4d \ ^{3}\mathrm{D}_{1}$\xspace & 3260.000 & 0.602~\cite{trautmann_quadrupole_transition} \\
        $5s4d \ ^{3}\mathrm{D}_{2}$\xspace & 3320.000 & 2.331~\cite{trautmann_quadrupole_transition} \\
        $5s4d \ ^{3}\mathrm{D}_{3}$\xspace & 3421.000 & 5.530~\cite{trautmann_quadrupole_transition} \\
        $5s4s \ ^{1}\mathrm{D}_{2}$\xspace & 5251.000 & 0.102~\cite{trautmann_quadrupole_transition} \\
        $5s6s \ ^{3}\mathrm{S}_{1}$\xspace & 14140.000 & 4.521~\cite{trautmann_quadrupole_transition} \\
        $5s5d \ ^{1}\mathrm{D}_{2}$\xspace & 19829.000 & 0.365~\cite{trautmann_quadrupole_transition} \\
        $5s5d \ ^{3}\mathrm{D}_{1}$\xspace & 20108.000 & 0.460~\cite{trautmann_quadrupole_transition} \\
        $5s5d \ ^{3}\mathrm{D}_{2}$\xspace & 20123.000 & 1.956~\cite{trautmann_quadrupole_transition} \\
        $5s5d \ ^{3}\mathrm{D}_{3}$\xspace & 20146.000 & 4.994~\cite{trautmann_quadrupole_transition} \\
        $5p^{2} \ ^{3}\mathrm{P}_{1}$\xspace & 20502.000 & 2.992~\cite{trautmann_quadrupole_transition} \\
        $5p^{2} \ ^{3}\mathrm{P}_{2}$\xspace & 20776.000 & 5.119~\cite{trautmann_quadrupole_transition} \\
        $5p^{2} \ ^{1}\mathrm{D}_{2}$\xspace & 22062.000 & 0.682~\cite{trautmann_quadrupole_transition} \\
        $5s7s \ ^{3}\mathrm{S}_{1}$\xspace & 22526.000 & 1.264~\cite{trautmann_quadrupole_transition} \\

\bottomrule
    \end{tabular}
    \label{tab:rdme_green}
\end{table}

\begin{table}[!h]
    \centering
    \caption{Relevant Wavelength and RDMEs for the Excited Manifold in \bluemotleak.}
    \begin{tabular}{ccc}
\toprule
        State $k$ & $\Delta E_{ki}$ (\si{\per\centi\meter}) & $\Bra{k} D \Ket{i}$ (a.u.) \\
\midrule
        $5s5p \ ^{3}\mathrm{P}_{1}$\xspace & 5645.351 & 0.186~\cite{safronovapriv} \\
        $5s5p \ ^{3}\mathrm{P}_{2}$\xspace & 5251.141 & 0.103~\cite{safronovapriv} \\
        $5s5p \ ^{1}\mathrm{P}_{1}$\xspace & 1548.767 & 1.926~\cite{safronovapriv} \\
        $4d5p \ ^{3}\mathrm{F}_{2}$\xspace & 13117.162 & 3.010~\cite{safronovapriv} \\
        $4d5p \ ^{3}\mathrm{F}_{3}$\xspace & 13440.029 & 0.227~\cite{safronovapriv} \\
        $4d5p \ ^{1}\mathrm{D}_{2}$\xspace & 13677.215 & 6.113~\cite{safronovapriv} \\
        $5s6p \ ^{3}\mathrm{P}_{1}$\xspace & 13718.626 & 0.671~\cite{safronovapriv} \\
        $5s6p \ ^{3}\mathrm{P}_{2}$\xspace & 13823.381 & 1.504~\cite{safronovapriv} \\
        $5s6p \ ^{1}\mathrm{P}_{1}$\xspace & 13948.713 & 2.257~\cite{safronovapriv} \\
        $4d5p \ ^{3}\mathrm{D}_{1}$\xspace & 16114.477 & 0.062~\cite{safronovapriv} \\
        $4d5p \ ^{3}\mathrm{D}_{2}$\xspace & 16232.053 & 0.256~\cite{safronovapriv} \\
        $4d5p \ ^{3}\mathrm{D}_{3}$\xspace & 16409.799 & 0.045~\cite{safronovapriv} \\
        $4d5p \ ^{3}\mathrm{P}_{1}$\xspace & 17153.041 & 0.060~\cite{safronovapriv} \\
        $4d5p \ ^{3}\mathrm{P}_{2}$\xspace & 17186.915 & 0.420~\cite{safronovapriv} \\
        $4d5p \ ^{1}\mathrm{F}_{3}$\xspace & 17858.068 & 0.442~\cite{safronovapriv} \\
        $5s4f \ ^{3}\mathrm{F}_{2}$\xspace & 18600.741 & 0.073~\cite{safronovapriv} \\
        $5s4f \ ^{3}\mathrm{F}_{3}$\xspace & 18602.713 & 0.145~\cite{safronovapriv} \\
        $5s7p \ ^{1}\mathrm{P}_{1}$\xspace & 18757.186 & 1.963~\cite{safronovapriv} \\
        $5s7f \ ^{1}\mathrm{F}_{3}$\xspace & 23513.152 & 1.431~\cite{NIST_ASD} \\
        $5s9p \ ^{1}\mathrm{P}_{1}$\xspace & 23184.739 & 1.243~\cite{NIST_ASD} \\
        $5s6f \ ^{1}\mathrm{F}_{3}$\xspace & 22696.277 & 1.883~\cite{NIST_ASD} \\
        $5s8p \ ^{1}\mathrm{P}_{1}$\xspace & 22318.903 & 1.591~\cite{NIST_ASD} \\
        $5s5f \ ^{1}\mathrm{F}_{3}$\xspace & 21375.167 & 2.593~\cite{NIST_ASD} \\
        $4d5p \ ^{1}\mathrm{P}_{1}$\xspace & 21028.305 & 1.829~\cite{NIST_ASD} \\
        $5s4f \ ^{1}\mathrm{F}_{3}$\xspace & 19394.729 & 3.576~\cite{NIST_ASD} \\

\bottomrule
    \end{tabular}
\end{table}

\subsection{Negative Sign of Tensor Light Shift}
\label{ssec:tensor_negative_sign}
Our light shift cancellation scheme is predicated on finding two wavelengths with comparable scalar polarizability (for yielding an isotropic trapping potential) and with opposite and comparable tensor polarizability (enabling the suppression of spin-dependent tensor light shifts). In this section we will illustrate the atomic structure conditions that enable us to estimate suitable wavelengths in \Sr. 

In Eq.~\eqref{eq:bare_pols} we can look closely at the bare tensor polarizability and isolate the angular momentum-specific term which depends on each dipole-allowed excited state transition summed over to yield the total tensor polarizability for a given wavelength, 
\begin{equation}
\begin{aligned}
(-1)^{F'+F} \bigg\{ & \begin{array}{ccc}
1 & 2 & 1 \\
F & F' & F 
\end{array} \bigg\} \\
= & (-1)^{-F + F'} \frac{\sqrt{\frac{2}{15}} \sqrt{\frac{\Gamma(2F+4)}{\Gamma(2F-1)}}}{(F+F')(F+F'+1)(F+F'+2)} \\
& \times\frac{1}{\Gamma(F-F'+2) \Gamma(-F+F'+2)}
\end{aligned}
\label{eq:tensor_qm_angular_momentum_phase_change_eq}
\end{equation}

In Eq.~\eqref{eq:tensor_qm_angular_momentum_phase_change_eq} we note that by construction $\Gamma$ functions will yield positive values, leaving $(-1)^{F'-F}$ as a term that changes tensor sign for a given excited state $\ket{n'F'}$. Let us define this phase term as $(-1)^{\Delta F}$ where $F'-F= \Delta F$. Now we take a closer look at the detuning contribution in the tensor polarizability, 
\begin{equation}
\frac{\omega_{n'F'nF}}{\omega_{n'F'nF} ^2
-\omega^2}
\label{eq:laser_detuning_scalar_tensor}
\end{equation}

Eq.~\eqref{eq:laser_detuning_scalar_tensor} introduces another factor in determining the tensor polarizability sign for a given frequency by $(\omega_{n'F'nF} -\omega)$. Finally, let us consider the reduced dipole matrix element (RDME) presented in Eq.~\eqref{eq:rdme_f_arno}. This term introduce a phase term dependent on the specific hyperfine state accounted for in the light shift calculation. As each excited hyperfine state $\ket{n'F'}$ yields a different phase factor, the overall tensor sign for a given electronic state is weighted by the RDME of a given hyperfine state, which in effect enhances the tensor sign in the calculation. 

\begin{figure}[!h]
\centering
    \includegraphics{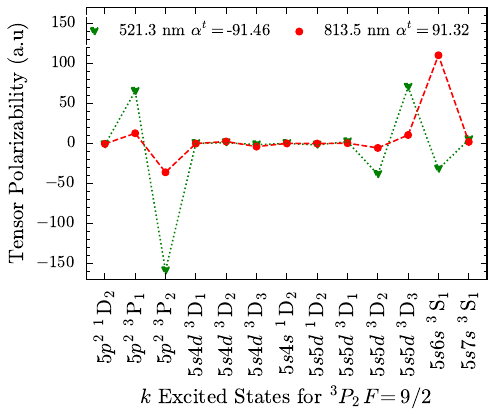}
        \caption{\textbf{Contributions of Excited Stats in \greenStatefull for Total Tensor Light Shift for 813.5\,nm and 521.3\,nm.} The dashed red line indicates the tensor light shift contributions for 813\,nm; each circle represents the total contribution for a given excited state. The dotted green line represents the tensor light shift contributions for 521\,nm; each triangle notes the corresponding contributions for each excited states for this wavelength.}
    \label{fig:negative_tensor}
\end{figure}

In Fig.~\ref{fig:negative_tensor} we present the breakdown of tensor light shift contribution for 813.5\,nm and 521.3\,nm. In the case of 813.5\,nm \greenState the excited state with the most impact for determining its tensor polarizability sign (and magnitude) is $5s6p \ ^\mathrm{3}\mathrm{S}_\mathrm{1}$ - as every other state is too far-detuned from 813.5\,nm to yield a meaningful impact on overall tensor sign. In contrast, for 521.3\,nm there are more excited states that impact the tensor light shift. In this case $5p^2\ ^\mathrm{3}\mathrm{P}_\mathrm{2}$ has the largest RDME (see Table.~\ref{tab:rdme_green}) which dominates the overall tensor light. 
\section{Rabi Frequencies in \greenState}
Given the time scale of 1\,ms we calculate the scale of the Rabi Frequencies in \greenState. We will use the fine structure estimate as a conservative figure of merit for viable coherent excitation rates. In Kl{\"u}sener et.\, al. the $A$ coefficient is reported to be 152\,\unit{\micro \s}~\cite{klusener}. To estimate the Rabi frequency they present a relationship between the decay rate $A$ and a reduced matrix element $\langle \gamma_eJ_e||H_{\mathrm{ph}}(pL)||\gamma_gJ_g\rangle$,
\begin{equation}
A = \dfrac{16\pi^2\alpha\nu_0}{2J_e+1}|\langle \gamma_eJ_e\|H_{\mathrm{ph}}(pL)\|\gamma_gJ_g\rangle|^2,
\label{eq:m2acoeff}
\end{equation}
with $\alpha$ the fine structure constant, $\nu_0$ is the frequency of the transition at 671.2\,nm (see Fig.~\ref{fig:overview}), $J_e =2$. Note: this matrix element is \emph{not} to be confused with the absolute value of the M2 $\langle ^1S_0\|\mathrm{M2}\|^3P_2\rangle $, which is  $22.6\mu_B$.

Furthermore, the Rabi frequency is defined as $\Omega = 2\pi E_0|V_{\mathrm{eg}}|/h$, where $h$ is the Planck constant and $V_{eg}$ is,
\begin{align}
\label{eq:VegSr}
\begin{split}
V_{eg} = &-\dfrac{\sqrt{\pi}ec}{2\pi\nu_0}\\
&\times
\langle \gamma_eJ_e\|H_{\mathrm{ph}}(0,2)\|\gamma_gJ_g\rangle \sum_{\lambda=\pm1}\lambda d^{2}_{0,\lambda}(\theta).
\end{split}
\end{align}
\noindent where $d^{2}_{0,\lambda}(\theta)$ is the Wigner d function and the measured value for $\theta = 14^{\circ}$. Moreover, both Eq.~\ref{eq:m2acoeff} and Eq.~\ref{eq:VegSr} and are in SI units. 

\begin{figure}[!h]
\centering
    \includegraphics{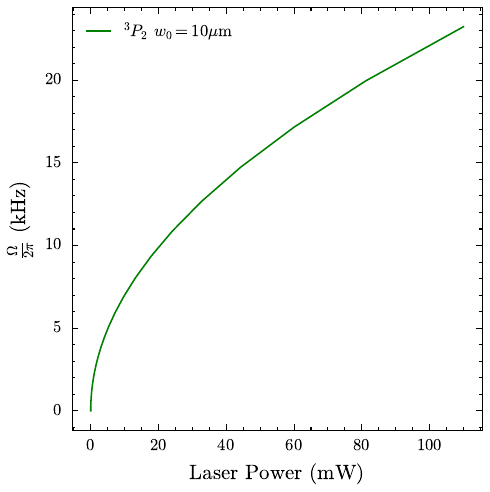}
        \caption{ \textbf{Estimated Rabi Rates in \Sr \greenState as a Function of Laser Power.} We assume a beam waist of 10 microns.} 
    \label{fig:m2_rabi_rates}
\end{figure}

In Fig.~\ref{fig:m2_rabi_rates} we present the estimated Rabi frequencies for a beam waist of 10 microns to address each atom in the array. These values indicate the viability of coherent excitation within the 1\,ms time scale we used in this paper to yield a state fidelity of 0.999. 
\subsection{Light Shifts for \bluemotleak near 521\,nm}
\label{sec:1d2_leaks}
In this section we present the light shift calculations near 521\,nm for \bluemotleak as this is a potential atom loss channel when operating imaging via broad-line cooling in \Sr. 

\begin{figure}[!h]
\centering
    \includegraphics{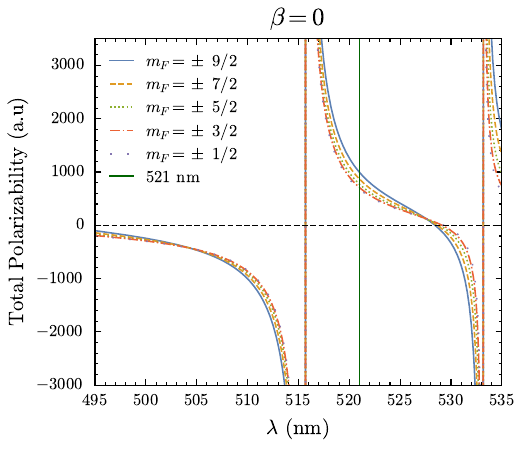}
        \caption{\textbf{Contributions of Excited Stats in \bluemotleak for Total Light Shift near 521\,nm.} The solid green indicates 521\,nm.}
    \label{fig:decay_path_pol}
\end{figure}

In Fig.~\ref{fig:decay_path_pol} we present the results for linearly polarized light. We present all nuclear spins in $F=9/2$ for \bluemotleak. At our chosen wavelength our trapping potential is trapping, mitigating atomic loss duirng imaging~\cite{caltech_thesis} experienced at 515\,nm due to \bluemotleak $\rightarrow$ $5s4f \ ^{1}\mathrm{F}_{3}$\xspace. Moreover, an atom in this state will experience a trapping potential at 813\,nm as well. 
\section{Tolerances in Bichromatic Tweezers for Yielding State Fidelity of 0.999 in \greenState in \Sr}
\label{ssec:more_hs}
\begin{figure}[!h]
\centering
    \includegraphics[width=1\columnwidth]{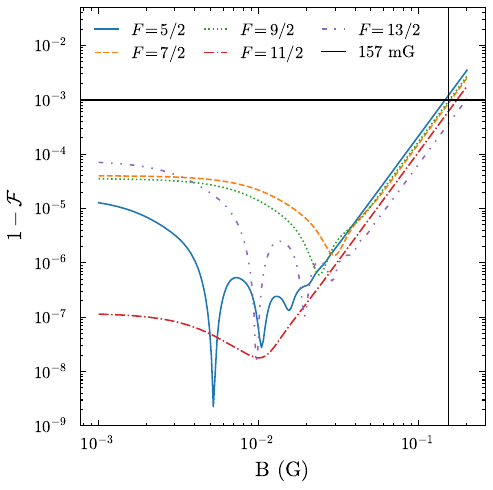}
            \caption{\textbf{State Fidelity as a Function of Magnetic Field Strength for 891.5\,nm and 518.0\,nm Bichromatic Tweezer at Tensor Magic Angle in \greenStatefull.} The solid blue line represents the fidelity for $F=5/2$. The dashed orange line represents the fidelity for $F=7/2$. The dotted green line represents the fidelity for $F=9/2$. The dash-dotted red line represents the fidelity for $F=11/2$. The loosely dash-dotted purple line represents the fidelity for $F=13/2$. All curves assume $x_0, \beta_0$ with no offsets or stochastic processes. Finally the solid black lines represent the level of magnetic field, 157\,mG, for yielding qudit fidelity of 0.999}
    \label{fig:hs_exact_fidelity_all_manifolds}
\end{figure}

For large-spin systems such as \Sr, and especially for qudit-based applications, in magnetically sensitive states such as \greenStatefull it is important to quantify impact of higher-order shifts induced by hyperfine mixing on qudit fidelity. We will use the Hilbert-Schmidt distance (see Eq.~\eqref{eq:fidelity_overlap_test}) to quantify this dephasing mechanism. In Fig.~\ref{fig:hs_exact_fidelity_all_manifolds} we present the estimation of the Hilbert-Schmidt overlap accounting for higher order Zeeman shifts introduced by hyperfine mixing in \greenStatefull for all hyperfine manifolds in this electronic state. We find that the largest magnetic field to yield our target fidelity of 0.999 for a 1\,ms hold time is at 157\,mG. We note that in this calculation we do not assume stochastic noise or systematic error offsets on the magic parameters $x_0, \beta_0$. As can be readily observed in Fig.~\ref{fig:hs_exact_fidelity_all_manifolds} all five of the hyperfine manifolds simultaneously experience a controlled mitigation of their hyperfine light shifts.
\subsection{Polarization Gradient-Induced Dephasing}
High numerical-aperture objectives~\cite{3p2_coherent_control,tao_qutrit, jeff_raman_debye,the_lunch_thesis} create spatial-dependent polarization gradients that introduce vector light shifts near the focus, introducing an additional dephasing mechanism. We can express this dephasing mechanism via the following Hamiltonian,

\begin{equation}
H_{\text{VLS}} = \frac{-1}{4} |E_0|^2 i \, \alpha_e^v(\omega)\frac{\mathbf{C}(r) \cdot \mathbf{ F}}{2F}
\label{eq:vector_ls_hamiltonian}
\end{equation}

\noindent where $\mathbf{C}(r) = \Im{ \mathbf{\epsilon} \cross \mathbf{\epsilon}^{*}}$ corresponds to the circularity of the electric field. Using the vector Debye integrals~\cite{Novotny_Hecht_2006} we compute the corresponding polarization vectors and estimate the scale of the induced light shifts.

\begin{figure}[!h]
\centering
    \includegraphics{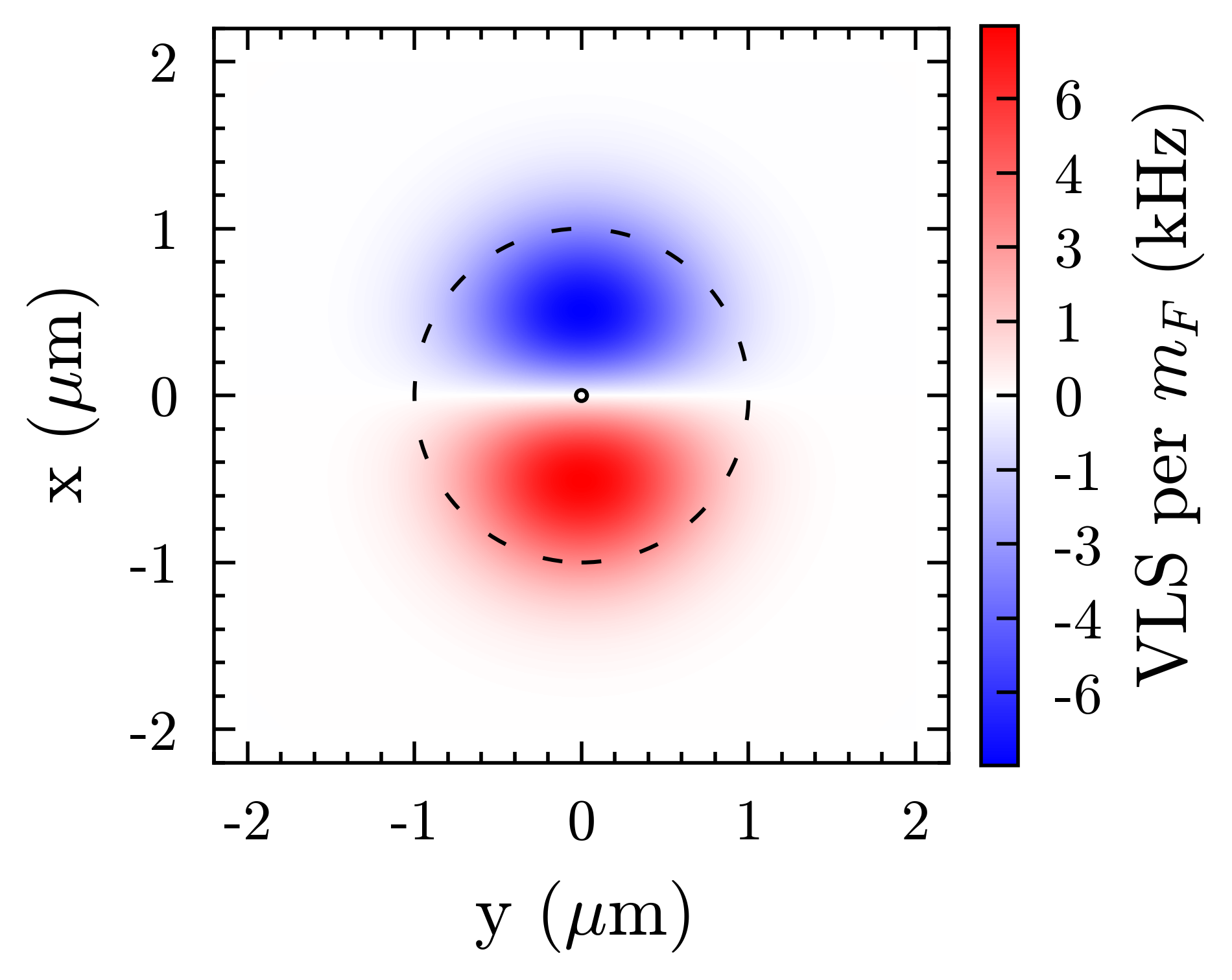}
        \caption{ \textbf{Polarization Gradient-Induced Vector Light Shifts in Bichromatic Tweezer using 891.5\,nm and 518.0\,nm for \Sr \greenState for $\beta=0$.} We present the vector light shifts per $m_F$ induced by circularity of the electric field near focus. The dashed circle represents the beam waist of 1 micron, while the solid circle represents the position uncertainty, 33\,nm, of a qudit in its motional ground state.} 
    \label{fig:debye_vector}
\end{figure}

In Fig.~\ref{fig:debye_vector} we present the total differential vector light shifts per nuclear spin sampled by the atom in its motional ground state at $x, y,$ at $z=0$. Given the relevant length scale and the symmetry of the vector light shifts, the atom would sample a light shift maximum of $\pm 1$\,\unit{\kilo \hertz} per $m_F$ at the edge of this length scale. This effect can be mitigated via dynamical decoupling sequences~\cite{klusener,sashas_paper}. Finally, we note that at the center of the tweezer there is effectively no effect due to vector light shifts, ensuring a cancellation of tensor light shifts in \greenState.

\section{Quadratic Zeeman Shift in \greenStatefull}
\label{appendix:quadratic_zeeman}
\subsection{Estimation of Quadratic Zeeman}
The magnetic sensitivity in \greenState introduces a dephasing mechanism via the quadratic Zeeman
shift~\cite{boyd_nuclear_spins}. In this section we will calculate this effect and its impact on our light shift cancellation scheme. First, let's define the Hamiltonian of interest,
\begin{multline}
H_{\text{total}}=H_Z+ H_{\text{hfs}}\\ 
H_{\text{hfs}} = A\vec{I}\cdot\vec{J}
+Q\frac{\frac{3}{2}\vec{I}\cdot\vec{J}(2\vec{I}\cdot\vec{J}+1)-IJ(I+1)(J+1)}{2IJ(2I-1)(2J-1)}
\label{eq:general-hamilonian}
\end{multline}
\noindent where $A$ and $Q$ are the magnetic dipole and quadrupole constants (for \greenStatefull $A = -212.765$\,MHz and $Q=67.215$\,MHz~\cite{3p2_hyperfine_heider}). Next, we define the Zeeman Hamiltonian $H_Z$ as,
\begin{equation}
    H_Z = (g_s S_z +g_l L_z - g_I I_z) \mu_B B
\end{equation}
\noindent where the $g_s \approx 2.00232$ and $g_l =1$ represent the electronic Land\'e $g$-factors; $g_I = -131.7712 \, \times 10^{-6} $ represents measured~\cite{florian_sr_rb} nuclear  Land\'e $g$-factor; $\mu_B$ is the Bohr magneton. We note that in this framework we set the magnetic field aligned with the $z$ axis.

Following the formalism presented on Lurio et al.~\cite{lurio_zeeman} we can write the matrix elements in our chosen basis $\ket{F, m_F}$,
\begin{widetext}
\begin{multline}
\langle S' L' J' F' m_F' | H_z | S L J F m_F \rangle = \delta_{S S'} \delta_{m_F m_F'} (-1)^{F' - m_F'} 
\begin{pmatrix} 
F' & 1 & F \\ 
-m_F' & 0 & m_F 
\end{pmatrix} \sqrt{(2F' + 1)(2F + 1)} \\
\times \Bigg[ \delta_{J J'} (-1)^{I + J + F + 1} 
\begin{Bmatrix} 
I & F' & J \\ 
F & I & 1 
\end{Bmatrix} \langle I || H_{z}^n || I \rangle 
+ (-1)^{I + J + F' + 1} 
\begin{Bmatrix} 
J' & F' & I \\ 
F & J & 1 
\end{Bmatrix} \langle J' || H_{z}^e || J \rangle \Bigg]
\label{eq:zeeman_matrix_element}
\end{multline}
\noindent where the electronic and nuclear reduced elements can be written~\cite{riegger_thesis} as,
\begin{multline}
\langle J' || H_{z}^e || J \rangle = \mu_B B \sqrt{(2J + 1)(2J' + 1)}
\times \left[ g_L (-1)^{L+S+J'+1} 
\begin{Bmatrix} 
L & J' & S \\ 
J & L & 1 
\end{Bmatrix} \sqrt{L(L + 1)(2L + 1)} \right. \\
\left. + g_S (-1)^{L+S+J+1} 
\begin{Bmatrix} 
S & J' & L \\ 
J & S & 1 
\end{Bmatrix} \sqrt{S(S + 1)(2S + 1)} \right]
\label{eq:electronic_rdme}
\end{multline}
\begin{equation}
    \langle I ||  H_{z}^n || I \rangle = -g_I \sqrt{I(I+1)(2I+1)} \mu_B B
\label{eq:nuclear_rdme}
\end{equation}
\end{widetext}
\noindent where $(\cdot)$ and $ \left \{ \cdot \right\}$ represent the $3j$ and $6j$ Wigner symbols. Furthermore, in our chosen basis the hyperfine interaction Hamiltonian is diagonal, yielding the following eigenvalues, 
\begin{equation}
    E_{HFS}(F) = \frac{A}{2} K + \frac{Q}{4} \frac{\frac{3}{2} K(K + 1) - 2I(I + 1)J(J + 1)}{I(2I - 1)J(2J - 1)}
\end{equation}
\begin{equation}
    K = F(F + 1) - I(I + 1) - J(J + 1)
\label{eq:hfs_energy}
\end{equation}

\begin{figure}[!h]
\centering
    \includegraphics{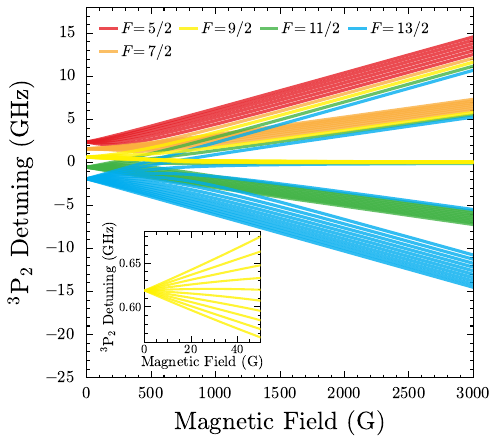}
        \caption{ \textbf{Breit-Rabi Diagram for \greenStatefull Manifold.} Inset shows the relevant hyperfine manifold of interest, \greenStatefull $F=9/2$, at the magnetic fields relevant for the calculation of the quadratic Zeeman shift.}
    \label{fig:breit_rabi_3p2}
\end{figure}

Armed with this framework we pursue the estimation of the quadratic Zeeman shift as follows: we will (1) build the basis states for the entire manifold (50 in total); (2) iterate through the basis set and apply the rules presented in Eq.~\eqref{eq:nuclear_rdme} and Eq.~\eqref{eq:electronic_rdme} to calculate the Zeeman matrix elements in Eq.~\eqref{eq:zeeman_matrix_element}; (3) estimate the corresponding hyperfine interaction matrix via Eq.~\eqref{eq:hfs_energy}. 

In Fig.~\ref{fig:breit_rabi_3p2} we present the Breit-Rabi diagram for \greenStatefull manifold in the $\ket{F,m_F}$ basis resulting from the full diagonalization of Eq.~\eqref{eq:general-hamilonian}. Using the data in the inset of Fig.~\ref{fig:breit_rabi_3p2}, we extract a quadratic Zeeman shift of the form:

\begin{equation}
\Delta^2_B(m_F, B) =  (188.7\,\mathrm{Hz/G^2}) \, m_F^2  \, \mathrm{B}^2
\label{eq:quad_zeeman_shift_eq}
\end{equation}

\subsection{Shift Sensitivity Evaluation in \greenStatefull}
\label{appendix:pt_notes}
For preserving qudit coherence the bichromatic tweezers needs to suppress spin dependent-shifts. For tweezer-induced light shifts introduced in an isolated manifold, this is achieved by appropriate choice of tweezer wavelengths and power ratio. However, for a manifold such as \greenStatefull, hyperfine mixing introduces additional dephasing mechanisms. In this section we will evaluate the shifts introduced by neighboring hyperfine manifolds via light shifts and Zeeman shifts through the lens of perturbation theory. 

First, let us present the Hamiltonian of interest, 
\begin{equation}
H = H_{\text{hfs}} + H_{z} + H_{\text{LS}}
\end{equation}

\noindent where $H_{\text{hfs}}$ represents the hyperfine interaction (see Eq.~\eqref{eq:general-hamilonian}). Now, we choose the $\ket{F, m_F}$ basis of the hyperfine interaction Hamiltonian and define the perturbation $V = H_{z} + H_{\text{LS}}$. This allows for the calculation of the first order correction for the bichromatic tweezer,
\begin{equation}
\begin{aligned}
\Delta_{F,m_F}^{(1)} &= \bra{F, m_F} H_z + H_{\text{LS}} \ket{F, m_F} \\
           &= \mu_B g_F B_z m_F + \frac{I_0}{2 \epsilon_0 c} \left[ \bar{\alpha}^{s}(\lambda_1, \lambda_2, x_0)  \right. \\
           &\quad \left. + \bar{\alpha}^{t}(\lambda_1, \lambda_2, x_0) \, G(\boldsymbol{\epsilon}, \hat{\mathbf{F}} ) \, \frac{1}{2F(2F-1)} \right]
\end{aligned}
\end{equation}
\noindent $I_0$ represents the nominal intensity of the optical tweezer set by $P_0$. Furthermore, we encapsulate the coupling of the internal degrees of freedom with light polarization as,
\begin{equation}
G(\boldsymbol{\epsilon}, \hat{\mathbf{F}} ) = 3 \left[
            \left( \boldsymbol{\epsilon}^{*} \cdot \hat{\mathbf{F}} \right) \left( \boldsymbol{\epsilon} \cdot \hat{\mathbf{F}} \right)
            + \left( \boldsymbol{\epsilon} \cdot \hat{\mathbf{F}} \right) \left( \boldsymbol{\epsilon}^{*} \cdot \hat{\mathbf{F}} \right)\right]  - 2 \hat{\mathbf{F}}^2
\end{equation}

Setting $x_0$ enables the suppression of differential tensor shifts leading to,
\begin{equation}
\Delta_{F,m_F}^{(1)} = \mu_B g_F B_z m_F +  \frac{I_0}{2 \epsilon_0 c} \bar{\alpha}^{s}(\lambda_1, \lambda_2, x_0) 
\end{equation}

Now let us calculate the corresponding second order shift,
\begin{equation}
\Delta_{F, m_F}^{(2)} = \sum_{F' \neq F, m_{F'}} \frac{\left| \bra{F', m_{F'}} H_z + H_{\text{LS}} \ket{F, m_F} \right|^2}{E_F^{(0)} - E_{F'}^{(0)}}
\end{equation}
\noindent where $E_F^{(0)}$ are represented by the hyperfine splittings for each corresponding manifold. Now we present the full expansion of this shift,
\begin{equation}
\begin{aligned}
\Delta_{F, m_F}^{(2)} &= \sum_{F' \neq F, m_{F'}} \frac{|\bra{F', m_{F'}} H_z \ket{F, m_F}|^2}{E_F^{(0)} - E_{F'}^{(0)}} \\
&+ \sum_{F' \neq F, m_{F'}} \frac{|\bra{F', m_{F'}} H_{\text{LS}} \ket{F, m_F}|^2}{E_F^{(0)} - E_{F'}^{(0)}} \\
&+ \sum_{F' \neq F, m_{F'}} \frac{ 2  \bra{F, m_F} H_z \ket{F', m_{F'}}}{E_F^{(0)} - E_{F'}^{(0)}} \\
&\quad \times  \bra{F', m_{F'}} H_{\text{LS}} \ket{F, m_F}
\end{aligned}
\label{eq:second_order_shift_overlap}
\end{equation}

At a glance we can identify 3 terms : 1) the quadratic Zeeman shift, 2) the quadratic light shift, and 3) the overlap shift between the magnetic and the light-matter interaction, leading to decoupling of the light-matter interaction from the Zeeman interaction, which we refer to as a magneto-optic shift~\cite{magneto_optic_shift}.

Now we turn to the estimation of the quadratic light shift which we calculate via the off-diagonal matrix elements~\cite{arnospaper,rosenbuschpaper},
\begin{equation}
\begin{aligned}
&\bra{F', m_{F'}} H_{\text{LS}} \ket{F, m_F} = \frac{I_0}{2 \epsilon_0 c} \sum_{K, q} (-1)^{K}(-1)^{q} \\
&\quad \times \{ \mathbf{\epsilon}^* \otimes \mathbf{\epsilon} \}_{q}^{K} (-1)^{F'-m_{F'}} 
\begin{pmatrix} 
F' & K & F \\ 
-m_{F'} & -q & m_F 
\end{pmatrix} 
\alpha_{nJF'F}^{(K)}(\omega)
\end{aligned}
\label{eq:spherical_ls_matrix_element}
\end{equation}
\noindent where $I_0$ represents the total Gaussian beam intensity. Now let us address the term $\{ \mathbf{\epsilon}^* \otimes \mathbf{\epsilon} \}_{q}^{K}$. This term represents the irreducible spherical tensor components of the light polarization for a given $K$ rank ($K=0,1,2$ corresponds to scalar, vector and tensor polarizability) and $q$ polarization degree of freedom~\cite{arnospaper},
\begin{equation}
\begin{split}
\{ \mathbf{\epsilon}^* \otimes \mathbf{\epsilon} \}_{q}^{K} = \sum_{\mu, \mu' = 0, \pm 1} & (-1)^{q+\mu'} \epsilon_{\mu} \epsilon_{-\mu'}^* \sqrt{2K+1} \\
& \times \begin{pmatrix} 
1 & K & 1 \\ 
\mu & -q & \mu' 
\end{pmatrix}
\end{split}
\end{equation}

\noindent where the reduced hyperfine structure polarizabilities can be written as~\cite{arnospaper,rosenbuschpaper},
\begin{equation}
\begin{aligned}
\alpha_{nJF' F}^{(K)}(\omega) &= (-1)^{J+I+F'+K} \sqrt{(2F + 1)(2F' + 1)} \\
&\quad \times 
\begin{Bmatrix} 
F' & K & F \\ 
J & I & J 
\end{Bmatrix} 
\alpha_{nJ}^{(K)}(\omega)
\end{aligned}
\end{equation}
\noindent represents the $\alpha_{nJ}^{(K)}(\omega)$ represents the reduced fine structure polarizability.

With this formalism, we can address the sensitivity set by light and Zeeman shifts. We have shown that our light shift cancellation scheme imposes a limit on light shifts (within the computational basis and those introduced by hyperfine mixing) through the power ratio, which in turn also suppresses this magneto-optic shift. Moreover, we underscore that this magneto-optic shift introduces $m_F^3$ shift in our system: $m_F$ from the Zeeman Hamiltonian and $m_F^2$ from the light shift Hamiltonian as encoded in the $3j$-symbol for each Hamiltonian (see Eq.~\eqref{eq:quad_zeeman_shift_eq} and Eq.~\eqref{eq:spherical_ls_matrix_element}). Furthermore, this dephasing mechanism is enhanced by \emph{both} energy scales, $I_0$ and $g_F \mu_B \mathrm{B}$, of this system. 

For the parameters presented in this paper (specifically for the 891.5\,nm and 518.0\,nm configuration) we find the scale of hyperfine mixing induced light shifts at the level of +8.8\,Hz, -8.4\,Hz. Thus, the fundamental dephasing mechanism in this system is set by the quadratic Zeeman shift, 
\begin{equation}
\Delta_{F, m_F}^{(2)}  = \sum_{F' \neq F, m_{F'}} \frac{|\bra{F', m_{F'}} H_z \ket{F, m_F}|^2}{E_F^{(0)} - E_{F'}^{(0)}} 
\end{equation}

To mitigate the effect of this dephasing mechanism we find that $\mathrm{B}$ $\lesssim 157$\,mG is able to maintain a state fidelity of 0.999. However, to retain the integrity of the fidelity plateau (allowing more tunability in the power ratios for specific angular offsets), $\mathrm{B}$ $\approx$ 100\,mG is preferred.

\section{Kramers--Heisenberg Formalism}
\label{sec:scattering_theory}
\subsection{Raman Scattering}
In this section we present the formalism for calculating Raman and Rayleigh scattering rates. We will follow the formalism laid out in the literature~\cite{lisdat_scattering, mike_thesis}. We will address two types of errors: leakage within the computational manifold and leakages to \clockState and \redState.

We start from the Kramers--Heisenberg formula  to calculate the scattering rate $\Gamma_{i \rightarrow f}$ of an atom experiencing a transition $i \rightarrow f$:
~\cite{mike_thesis, lisdat_scattering}
\begin{equation}
	\Gamma_{i \rightarrow f} = \frac{I {{\omega_{sc}^3}}}{\left(4\pi \mathbf{\epsilon}_{0} \right)^2 c^4 \hbar^3} \frac{8\pi}{3}  \left| \mathbf{D}(\omega)\right|^2
 	\label{eq:sc_rate}
\end{equation}

\noindent where $\left| \mathbf{D}(\omega)\right|^2$ is defined as the net induced dipole  for the scattered radiation~\cite{mike_thesis}. Induced pumping mechanisms leading to $i \rightarrow j$ correspond to Raman scattering processes, which lead to leakage and depolarization errors in qudits. In contrast, for pumping mechanisms inducing $i \rightarrow i$ events correspond to traditional Rayleigh scattering processes, which lead to dephasing errors. 

We can cast the net induced dipole in terms of the polarization degrees of freedom $q$ as,

\begin{equation}
    \left| \mathbf{D}(\omega)\right|^2 = \sum_{q=-1}^{1} \left| D_q^{(i \rightarrow f)} (\omega) \right|^2
\end{equation}

Returning to our calculation we will frame Eq.~\eqref{eq:sc_rate} as, 
\begin{equation}
	\Gamma_{i \rightarrow f} = \frac{I {\omega_{sc}^3}}{\left(4\pi \mathbf{\epsilon}_{0} \right)^2 c^4 \hbar^3} \frac{8\pi}{3} \sum_{q=-1}^{1} \left| D_q^{(i \rightarrow f)}(\omega) \right|^2
\end{equation}

The dipole operator for a polarization vector $\hat{\epsilon}$ can be expressed as:
\[
d(\epsilon)= \vec{\mathbf{r}} \cdot \vec{\boldsymbol{\epsilon}}
\]
and for the scattered polarization as
\[
d(\epsilon_{\lambda}) = \vec{\mathbf{r}} \cdot \vec{\boldsymbol{\epsilon_{\lambda}}^{*}}
\]

\noindent where for simplicity we will cast $d$ as the dipole introduced by the incoming radiation and $d_{\lambda}$ as the dipole of the scattered photons.

Thus we can write the net dipole as,
\begin{multline}
D_q^{(i \rightarrow f)}(\omega) = \sum_k \Bigg(
\frac{
    \langle f |  d_{\lambda,q} | k \rangle 
    \langle k | d_q | i \rangle
}{\omega_k - \omega}
\\
+
\frac{
    \langle f |d_q | k \rangle
    \langle k | d_{\lambda,q} | i \rangle
}{\omega_k + \omega_{sc}}
\Bigg)
\label{eq:sc_amplitude}
\end{multline}
\noindent where we have casted away the vector nature of the operator encoded in the matrix element. Furthermore we encode the polarization state (in the spherical basis) via $q$ to both incoming and scattered photons~\cite{lisdat_scattering}.

\subsubsection{$\pi$-polarized light}
For this calculation we will follow the traditional framework in which we assume the tweezer light to be $\pi$ polarized (and that is aligned with the quantization axis, $\beta=0$); thus $d_q \rightarrow d_0$. Moreover, we can express each component of the net induced dipole as a function of the electric dipole $d_q$ driving this process,
\begin{equation}
	\begin{split}
		D_q^{(i \rightarrow f)}(\omega) = \sum_{k} &\left( \matrixelement{f}{ d_{\lambda,q}}{k} \frac{\matrixelement{k}{d_{0}}{i}} {\omega_{ki} - \omega} + \right. \\
		                                   &\left. \matrixelement{f}{d_{0}}{k} \frac{\matrixelement{k}{ d_{\lambda,q}}{i}} {\omega_{ki} + \omega_{sc}}  \ \right)
	\end{split}
\end{equation}
\noindent where we sum over all allowed $k$ excited states. Furthermore, we will define $\omega_{sc}$, 
\begin{eqnarray}
    \omega_{sc} = \omega - \omega_{fi} \\
    \omega_{fi} = \omega_f - \omega_i
\end{eqnarray}

Furthermore, we will cast this formula in terms of scattering amplitudes~\cite{era_joanna,era_uys}
\begin{multline}
    A^{i\rightarrow
j}_{k,q}(\omega) = \left( \matrixelement{f}{ d_{\lambda,q}}{k} \frac{\matrixelement{k}{d_{0}}{i}} {\omega_{ki} - \omega} + \right. \\
		                                   \left. \matrixelement{f}{d_{0}}{k} \frac{\matrixelement{k}{ d_{\lambda,q}}{i}} {\omega_{ki} + \omega_{sc}} \ \right)
\end{multline}

$i$ represents $\ket{F_1, m_1}$ and $f$ represents $\ket{F_2, m_2}$. For most of our analysis $\ket{F_2, m_2}$  represents states in \metastablemanifold and $\ket{F_1,m_1}$ represents states in \greenState. For the purposes of this calculation $F_1 =$ \greenState $F=9/2$ and $F_2$ can be presented by either \redState or \clockState. Using the \wet,
\begin{equation}
\begin{split}
        \matrixelement{F_2, m_2}{d_{\lambda,q}}{F', m'} \matrixelement{F',m'}{d_{0}}{F_1,m_1}  = (-1)^{F_2 -m_2} \times \\ \bigg(\begin{array}{ccc}
F_2 &1 &F' \\
-m_2 & q & m'
\end{array}\bigg) \matrixelement{F_2}{|d|}{F'} \times \\  (-1)^{F' -m'} \times\bigg(\begin{array}{ccc}
F' &1 &F_1 \\
-m' & q=0 & m_1
\end{array}\bigg) \matrixelement{F'}{|d|}{F_1}
\end{split}
\end{equation}

Using the fact~\cite{LeKien2012DynamicalPO,walraven_notes,budker_book_ls},
\begin{equation}
    \matrixelement{F_2}{|d|}{F'} = (-1)^{F_2 - F'} \matrixelement{F'}{|d|}{F_2}^{*}
\end{equation}
\noindent We note that under that under this framework our RDME is more symmetric under exchange~\cite{caltech_thesis}, 
\begin{equation}
    \matrixelement{F'}{|d|}{F_2}^{*} = \matrixelement{F'}{|d|}{F_2} 
\end{equation}
We can decompose the first matrix elements,
\begin{equation}
\begin{split}
        \matrixelement{F_2, m_2}{d_{\lambda,q}}{F', m'} \matrixelement{F',m'}{d_{0}}{F_1,m_1}  = (-1)^{2 F_2 -m_2 -m'} \\ \bigg(\begin{array}{ccc}
F_2 &1 &F' \\
-m_2 & q & m'
\end{array}\bigg) \bigg(\begin{array}{ccc}
F' &1 &F_1 \\
-m' & q=0 & m_1
\end{array}\bigg) \matrixelement{F'}{|d|}{F_2} \matrixelement{F'}{|d|}{F_1}
\end{split}
\end{equation}

\noindent where the indexed $F_i$ indicates the appropriate state in which to obtain the reduced dipole matrix elements (RDMEs). Namely, there are two different RDMEs from each manifold considered for a given scattering for the process $F_1 \rightarrow F'$, where $F'$ is a virtual state, with a RDME accounting for that transition, while the secondary process $F' \rightarrow F_2$ to reach the final state requires a different RDME. Similarly, we can compute the second term in the net induced dipole, 
\begin{equation}
\begin{split}
         \matrixelement{F_2, m_2}{d_{0}}{F',m'} \matrixelement{F',m'}{d_{\lambda,q}}{F_1,m_1}  = (-1)^{F_2 -m_2} \\ \bigg(\begin{array}{ccc}
F_2 &1 &F' \\
-m_2 & q=0 & m'
\end{array}\bigg) \matrixelement{F_2}{|d|}{F'} \times \\  (-1)^{F' -m'} \times\bigg(\begin{array}{ccc}
F' &1 &F_1 \\
-m' & q & m_1
\end{array}\bigg) \matrixelement{F'}{|d|}{F_1}
\end{split}
\end{equation}
Yielding, 
\begin{equation}
\begin{split}
        \matrixelement{F_2, m_2}{d_{\lambda,q}}{F', m'} \matrixelement{F',m'}{d_{0}}{F_1,m_1}  = (-1)^{2 F_2 -m_2 -m'} \\ \bigg(\begin{array}{ccc}
F_2 &1 &F' \\
-m_2 & q=0 & m'
\end{array}\bigg)\bigg(\begin{array}{ccc}
F' &1 &F_1 \\
-m' & q & m_1
\end{array}\bigg) \matrixelement{F'}{|d|}{F_2}^{*} \matrixelement{F'}{|d|}{F_1}
\end{split}
\end{equation}

Thus we cast the general formula for calculating scattering amplitudes,
\begin{multline}
		A^{i\rightarrow
j}_{k,q}(\omega) = (-1)^{2 F_2 -m_2 -m'}\matrixelement{F'}{|d|}{F_2} \matrixelement{F'}{|d|}{F_1}  \\ \Bigg[\frac{1} {\omega_{ki} - \omega} \bigg(\begin{array}{ccc}
F_2 &1 &F' \\
-m_2 & q & m'
\end{array}\bigg) \bigg(\begin{array}{ccc}
F' &1 &F_1 \\
-m' & q=0 & m_1
\end{array}\bigg) \\ + 
		                                 \frac{1} {\omega_{ki} + \omega_{sc}} \bigg(\begin{array}{ccc}
F_2 &1 &F' \\
-m_2 & q=0 & m'
\end{array}\bigg)\bigg(\begin{array}{ccc}
F' &1 &F_1 \\
-m' & q & m_1
\end{array}\bigg) \Bigg]    
\end{multline}

Thus we can write Eq.~\eqref{eq:sc_amplitude} in terms of scattering amplitudes, 
\begin{equation}
		D_q^{(i \rightarrow f)}(\omega) = \sum_{k} A_{k,q}^{i \rightarrow j}(\omega) 
\end{equation}
Leading to a compressed Kramers--Heisenberg formula, 
\begin{equation}
	\Gamma_{i \rightarrow f} = \frac{I {\omega_{sc}^3}}{\left(4\pi \mathbf{\epsilon}_{0} \right)^2 c^4 \hbar^3} \frac{8\pi}{3}  \sum_{q=-1}^{1} \left| \sum_{k} A_{k,q}^{i \rightarrow j}(\omega) \right|^2
 \label{eq:compressed_kh}
\end{equation}

\subsubsection{Arbitrary Linearly Polarized Light}
\label{ssec:arbitrary_pi_light}
To capture the appropriate scattering rates for our tweezer operating at the magic angle let us cast the polarization vector parametrized by $\beta$ presented earlier in this paper in the spherical basis as,
\begin{equation}
\boldsymbol{\epsilon} (\beta)
= -\frac{\sin\beta}{\sqrt{2}}\, \hat{\mathbf{e}}_{+1}
+ \cos\beta\, \hat{\mathbf{e}}_0
+ \frac{\sin\beta}{\sqrt{2}}\,\hat{\mathbf{e}}_{-1}
\end{equation}

Given our choice of arbitrary linear polarization we can write the incoming photon dipole moment as~\cite{arb_dipole_pol},
\begin{equation}
 \boldsymbol{d}_{\lambda,q}(\beta) = -\frac{\sin\beta}{\sqrt{2}}\, \hat{\mathbf{d}}_{+1} + \cos(\beta)\,\hat{\mathbf{d}}_{0} + \frac{\sin\beta}{\sqrt{2}}\, \hat{\mathbf{d}}_{-1}
\end{equation}
Thus we yield a modified scattering amplitude,

\begin{widetext}
\begin{multline}
A^{i \rightarrow j}_{k,q}(\beta,\omega) = 
(-1)^{2 F_2 - m_2 - m'} 
\langle F' \| d \| F_2 \rangle
\langle F' \| d \| F_1 \rangle
\\
\times \Bigg\{
\frac{1}{\omega_{ki} - \omega}
\Bigg[
\left(
\begin{array}{ccc}
F_2 & 1 & F' \\
-m_2 & q & m'
\end{array}
\right)
\left(
\begin{array}{ccc}
F' & 1 & F_1 \\
-m' & q=0 & m_1
\end{array}
\right)
\cos\beta
+
\left(
\begin{array}{ccc}
F_2 & 1 & F' \\
-m_2 & q & m'
\end{array}
\right)
\left(
\begin{array}{ccc}
F' & 1 & F_1 \\
-m' & q=-1 & m_1
\end{array}
\right)
\frac{\sin\beta}{\sqrt{2}}
\\
-
\left(
\begin{array}{ccc}
F_2 & 1 & F' \\
-m_2 & q & m'
\end{array}
\right)
\left(
\begin{array}{ccc}
F' & 1 & F_1 \\
-m' & q=1 & m_1
\end{array}
\right)
\frac{\sin\beta}{\sqrt{2}}
\Bigg]
\\
+
\frac{1}{\omega_{ki} + \omega_{sc}}
\Bigg[
\left(
\begin{array}{ccc}
F_2 & 1 & F' \\
-m_2 & q=0 & m'
\end{array}
\right)
\left(
\begin{array}{ccc}
F' & 1 & F_1 \\
-m' & q & m_1
\end{array}
\right)
\cos\beta
+
\left(
\begin{array}{ccc}
F_2 & 1 & F' \\
-m_2 & q=-1 & m'
\end{array}
\right)
\left(
\begin{array}{ccc}
F' & 1 & F_1 \\
-m' & q & m_1
\end{array}
\right)
\frac{\sin\beta}{\sqrt{2}}
\\
-
\left(
\begin{array}{ccc}
F_2 & 1 & F' \\
-m_2 & q=1 & m'
\end{array}
\right)
\left(
\begin{array}{ccc}
F' & 1 & F_1 \\
-m' & q & m_1
\end{array}
\right)
\frac{\sin\beta}{\sqrt{2}}
\Bigg]
\Bigg\}
\end{multline}
\end{widetext}

We are interested in yielding a general framework in which to calculate Raman scattering errors of the form $\ket{^3\mathrm{P}_2, F, m_F} \rightarrow \ket{^3\mathrm{P}_J, F', m_{F'}}$ where \metastablemanifold corresponds to \redState, \clockState and \greenState. This framework will allow us to estimate scattering rates for leakages into other metastable states as well within \greenState. 

\subsection{Rayleigh Decoherence via Differential Scattering Amplitudes between $m_F$}
In addition to Raman scattering, far-detuned light fields can induce Rayleigh scattering. The literature defines elastic Rayleigh scattering as~\cite{era_uys}, 
\begin{equation}
\Gamma_{el}^{i,j}(\omega)= \frac{I {\omega}^3}{\left(4\pi \mathbf{\epsilon}_{0} \right)^2 c^4 \hbar^3} \frac{8\pi}{3} \sum_{q} \left(\sum_{k}A^{i\rightarrow
i}_{k,q}(\omega)-A^{j\rightarrow
j}_{k,q}(\omega)\right)^2
\end{equation}
Furthermore, let's define effective differential scattering amplitude between nuclear spins, 
\begin{equation}
\mathbf{\epsilon}(\omega) =\sum_{q}\Bigg(\sum_{k} A_{k,q}^{i,i}(\omega) - \sum_{k}A_{k,q}^{j,j}(\omega) \Bigg)
\end{equation} 
Thus we can see that dephasing errors scale quadratically~\cite{era_uys},
\begin{equation}
\Gamma_{el}^{i,j}(\omega)= \frac{I {\omega}^3}{\left(4\pi \mathbf{\epsilon}_{0} \right)^2 c^4 \hbar^3} \frac{8\pi}{3} \Bigg[\mathbf{\epsilon}(\omega)\Bigg]^2
\end{equation}

For this form of error the difference in phase between the scattering amplitudes of $i,j$ nuclear spins lead to constructive interference driving dephasing. For clarity, we will expand this expression (and adding $\beta$ as a parameter) to,
\begin{widetext}
\begin{equation}
    \Gamma_{el}^{\, i,j}(\beta,\omega)= \frac{I {\omega}^3}{\left(4\pi \mathbf{\epsilon}_{0} \right)^2 c^4 \hbar^3} \frac{8\pi}{3} \sum_{q} \Bigg[{\left( \sum_{k} A_{k,q}^{i \rightarrow i}(\beta,\omega) \right)^2} +\\ {\left(  \sum_{k} A_{k,q}^{j \rightarrow j}(\beta,\omega) \right)^2} - {2 \sum_{k} A_{k,q}^{i \rightarrow i}(\beta,\omega) \ A_{k,q}^{j \rightarrow j}(\beta,\omega)} \Bigg]
\label{eq:era_full}
\end{equation}
\end{widetext}
The first two terms represent standard Rayleigh scattering for $i,j$ nuclear spins in the qudit. However, the last term represents the overlap between the scattering amplitudes of the different nuclear spins, introducing a decoherence mechanism driving depolarization errors.

For qubit-systems, characterizing dephasing via this formalism is straight-forward. For qudits we need to expand this formalism by introducing $\mathcal{E}$, a matrix representing the elastic Rayleigh scattering in the SU(10) manifold.
\begin{widetext}
\begin{equation}
\begin{aligned}
    \mathcal{E}=  \begin{pmatrix}
0 & \Gamma_{el}^{-9/2,-7/2} & \Gamma_{el}^{-9/2,-5/2}  & \Gamma_{el}^{-9/2,-3/2} & \cdots & \Gamma_{el}^{-9/2,7/2} & \Gamma_{el}^{-9/2,9/2} \\
\Gamma_{el}^{-7/2,-9/2} & 0 & \Gamma_{el}^{-7/2,-5/2}  & \Gamma_{el}^{-7/2,-3/2} & \cdots &\vdots & \vdots \\
\Gamma_{el}^{-5/2,-9/2} & \Gamma_{el}^{-5/2,-7/2}  &  0 & \ddots & \cdots & \vdots & \vdots \\
\Gamma_{el}^{-3/2,-9/2} & \Gamma_{el}^{-3/2,-7/2} & \ddots &  \ddots & \ddots &\vdots &\vdots \\
\vdots & \vdots & \Gamma_{el}^{i,j} & \ddots & 0 & \vdots & \vdots \\
\Gamma_{el}^{7/2,-9/2} & \Gamma_{el}^{7/2,-7/2} & \vdots &  \cdots & \cdots & 0 & \Gamma_{el}^{7/2,9/2}\\
\Gamma_{el}^{9/2,-9/2} & \Gamma_{el}^{9/2,-7/2} & \Gamma_{el}^{9/2,-5/2} & \cdots & \Gamma_{el}^{9/2,5/2} &\Gamma_{el}^{9/2,7/2}  & 0
\end{pmatrix}
\end{aligned}
\label{eq:eva_matrix}
\end{equation}
\end{widetext}

From $\mathcal{E}$ we define the effective Rayleigh scattering rate for a nuclear spin as,
\begin{equation}
        \Gamma_{\text{eff}}(j) = \frac{1}{n}\sum_{i=1}^{n} \mathcal{E}_{ij}
\label{eq:effective_eva}
\end{equation}
In Eq.~\eqref{eq:effective_eva} $j$ is the column index representing nuclear spin.

\end{document}